\documentclass[aps,pra,reprint,superscriptaddress]{revtex4-1}
\usepackage{filecontents}
\usepackage{amssymb}
\usepackage{graphicx}
\usepackage{enumitem}
\usepackage{amsmath}
\newcommand{\be}{\begin{equation}}
\newcommand{\ee}{\end{equation}}
\newcommand{\ba}{\begin{eqnarray}}
\newcommand{\ea}{\end{eqnarray}}

\newcommand{\RN}[1]{%
  \textup{\uppercase\expandafter{\romannumeral#1}}%
}

\begin{document}

\title{Phase Diagram of Quantum Hall Breakdown and Non-linear Phenomena for InGaAs/InP Quantum Wells}
\author{V. Yu}
\affiliation{Department of Physics, McGill University, Montr\'eal, QC, Canada, H3A 2T8}
\affiliation{Emerging Technology Division, National Research Council of Canada, M50, 1200 Montreal Road, Ottawa, Ontario K1A 0R6, Canada}

\author{M. Hilke}
\email{hilke@physics.mcgill.ca}
\affiliation{Department of Physics, McGill University, Montr\'eal, QC, Canada, H3A 2T8}

\author{P. J. Poole}

\affiliation{Emerging Technology Division, National Research Council of Canada, M50, 1200 Montreal Road, Ottawa, Ontario K1A 0R6, Canada}

\author{S. Studenikin}

\affiliation{Emerging Technology Division, National Research Council of Canada, M50, 1200 Montreal Road, Ottawa, Ontario K1A 0R6, Canada}

\author{D. G. Austing}
\email{guy.austing@nrc-cnrc.gc.ca}
\affiliation{Department of Physics, McGill University, Montr\'eal, QC, Canada, H3A 2T8}
\affiliation{Emerging Technology Division, National Research Council of Canada, M50, 1200 Montreal Road, Ottawa, Ontario K1A 0R6, Canada}

\begin{abstract}
We investigate non-linear magneto-transport in a Hall bar device made from a strained InGaAs/InP quantum well: a material system with attractive spintronic properties. From extensive maps of the longitudinal differential resistance (r$_{xx}$) as a function of current and magnetic (B-) field phase diagrams are generated for quantum Hall breakdown in the strong quantum Hall regime reaching filling factor $\nu$=1. By careful illumination the electron sheet density (n) is incremented in small steps and this provides insight into how the transport characteristics evolve with n. We explore in depth the energetics of integer quantum Hall breakdown and provide a simple picture for the principal features in the r$_{xx}$ maps. A simple tunneling model that captures a number of the characteristic features is introduced. Parameters such as critical Hall electric fields and the exchange-enhanced g-factors for odd-filling factors including $\nu$=1 are extracted. A detailed examination is made of the B-field dependence of the critical current as determined by two different methods and compiled for different values of n. A simple rescaling procedure that allows the critical current data points obtained from r$_{xx}$ maxima for even-filling to collapse on to a single curve is demonstrated. Exchange-enhanced g-factors for odd-filling are extracted from the compiled data and are compared to those determined by conventional thermal activation measurements. The exchange-enhanced g-factor is found to increase with decreasing n.
\end{abstract}

\maketitle


\section{Introduction}\label{sec-intro}

The quantum Hall effect occurs in a two-dimensional electron gas (2DEG) confined in a quantum well (QW) when subjected to a high perpendicular magnetic field~\cite{vKLIT2017}. By passing a sufficiently high current through a Hall bar, the quantum Hall effect can be destroyed~\cite{EBERT1983}. Current-induced quantum Hall breakdown is a valuable tool to access non-linear magneto-transport phenomena in a 2DEG (see the extensive review in Ref.~\onlinecite{NACHT1999} and references therein). Some examples of the rich variety of non-linear phenomena that can be accessed in a strong magnetic field at and beyond current-induced quantum Hall breakdown include: hysteresis arising from dynamic nuclear polarization through the interplay of electron and nuclear spins via the hyperfine interaction near integer odd-filling factors~\cite{KAWA2007, KAWA2011}; cyclotron emission at terahertz frequency arising from non-equilibrium electron distribution in Landau levels~\cite{IKUSHIMA2006, IKUSHIMA2007}; electric instability leading to resistance fluctuations and negative differential resistance~\cite{EAVES2001A, EAVES2001B, CHEN2009, LEE2015}; and re-entrant quantum Hall states of the second Landau level ~\cite{BAER2015, ROSS2016}.

The above mentioned phenomena are typically observed in widely studied GaAs/AlGaAs hetero-structure 2DEGs with high mobility. Here we report on the general characteristics of quantum Hall breakdown in an InGaAs/InP QW Hall bar device. Such indium-based QWs are of current interest for spintronic applications. As compared to widely employed GaAs/AlGaAs QWs, InGaAs/InP QWs offer larger electron g-factors, stronger spin-orbit coupling, and the nuclear spin for indium (9/2) is larger than for gallium (3/2)~\cite{DATTA1990, NITTA1997, ZUTIC2004, HIRA2009, KOHDA2012, SHAB2016, KOHDA2017}. We examine in detail two-dimensional maps of the differential resistance as a function of current and magnetic field. These maps may be regarded as phase diagrams and provide a wealth of information about quantum Hall breakdown (including the conditions for, and the energetics of, quantum Hall breakdown), and a number of other phenomena for sheet current densities up to $\sim$1~A/m. We describe how quantum Hall breakdown and non-linear phenomena evolve systematically not only with increasing magnetic field (B) up to 9~T (equivalently decreasing filling factor $\nu$) but also with increasing electron sheet density from 1.6$\times$10$^{11}$~cm$^{-2}$ up to 3.9$\times$10$^{11}$~cm$^{-2}$. The capability to increment the electron density in small steps by careful illumination was particularly valuable and enabled us to compile sufficient data to examine, for example, trends in electronic g-factors, and to compare different methods by which the breakdown current can be extracted. In contrast to an earlier study of an InGaAs/InP QW in Ref.~\onlinecite{STUD2012}, here we access the strong quantum Hall regime ($\nu$=1). We note there are some reports of current-induced breakdown in other indium-based QW systems where $\nu$=1 is also attained (see Ref.~\onlinecite{NACHT2000} for InGaAs/InAlAs QWs and Ref.~\onlinecite{ALEXANDAR2012} for InSb/AlInSb QWs).

The outline of this paper is as follows. Section II describes details of the InGaAs/InP hetero-structure and the Hall bar investigated, and transport characteristics of the 2DEG as a function of the electron density. Section III explains how differential resistance colormaps are generated, and outlines the principal features revealed. Section IV describes how the breakdown characteristics evolve with electron density. Section V provides a simple picture for the principal features in the differential resistance colormaps and introduces a simple tunneling model that captures a number of characteristic features. Section VI describes parameters that can be extracted from the transport diamonds in the strong quantum Hall regime: principally the critical Hall electric fields for $\nu$=1 and $\nu$=2 and an estimate of the g-factor for $\nu$=1. Section VII examines in detail the B-field dependence of the critical breakdown current as determined by two different methods compiled for different electron densities. g-factors for odd-filling factors  extracted from the compiled data are compared to those determined by conventional thermal activation measurements. Lastly, Section VIII discusses a curious and as yet unexplained zero current anomaly readily visible in the experimental data over a wide B-field range.

\section{EXPERIMENTAL DETAILS AND 2DEG CHARACTERISTICS}

The 2DEG investigated is formed in the QW region of a strained In$_{0.76}$Ga$_{0.24}$As/In$_{0.53}$Ga$_{0.47}$As/InP hetero-structure grown by chemical beam epitaxy. The conducting channel in the QW is an undoped 10~nm In$_{0.76}$Ga$_{0.24}$As layer. As compared to lattice matched In$_{0.53}$Ga$_{0.47}$As/InP QWs, the 0.76 indium fraction is intended to reduce the effective mass and maximize the mobility at the cost of introducing strain~\cite{HARD1993}. Because of the mobility enhancement, In$_x$Ga$_{1-x}$As QWs with In-fraction x of $\sim$0.75 have attracted attention over the years~\cite{CHIN1990, HARD1992, RAMVALL1998, GOZU2001, STUD2003, DESRAT2004, ERCOLANI2008, HERZOG2015, CHEN2015}. Other details of the growth and hetero-structure are given in Refs.~\onlinecite{STUD2012, VICTORTHESIS}. All the transport measurements described in this report are for a 15~$\mu$m wide Hall bar prepared by standard optical lithography and wet etching~\cite{VICTORTHESIS}. Three pairs of potential probes separated by 50~$\mu$m are positioned along the Hall bar and the potential probes have width 7.5~$\mu$m where they join the Hall bar. The Hall bar is $\sim$200~$\mu$m long and opens out at either end to avoid sharp corners on the entry to the source and drain regions. The Hall bar device design and dimensions were motivated by the work in Refs.~\onlinecite{KAWA2007, KAWA2011}. Soldered indium beads that are subsequently annealed at 350$^{\circ}$C in vacuum for 30~mins form Ohmic contacts. The device is maintained at the base temperature of a top-loading $^3$He Janis cryostat equipped with a 9~T superconducting magnet.

\begin{figure}[!h]
	\centering
	\includegraphics[width=0.4\textwidth]{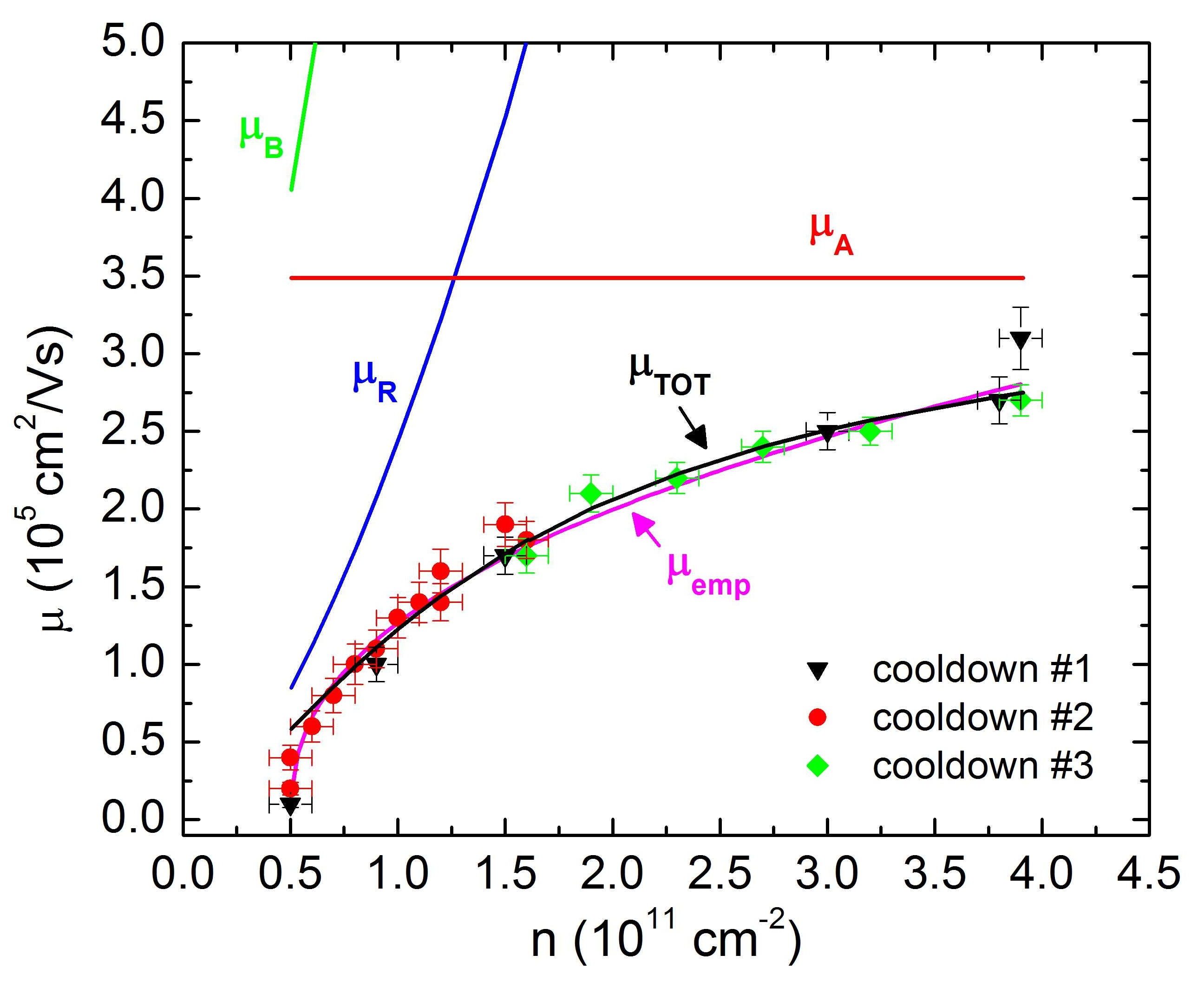}\\
	\caption{0.3~K values of electron density n and mobility $\mu$ attained by illumination over three separate cooldowns \#1-\#3. Estimated uncertainties in n and $\mu$ are included for each data point. Empirically, over the density range accessible, we find the mobility follows a power-law dependence $\mu_{emp}{\sim}(\mbox{n} - \mbox{n}_0)^p$ where n$_0$=0.50$\pm$0.02$\times$10$^{11}$~cm$^{-2}$ and p=0.42$\pm$0.02 (see dashed magneta trace).  Following the model in Ref.~\onlinecite{DASSARMA2014} we have also fitted our data for ionized impurity scattering and short range alloy disorder scattering. Included in the figure are the calculated mobility versus density traces for background ionized impurity scattering ($\mu_B$: green trace), remote ionized impurity scattering ($\mu_R$: blue trace), short range alloy disorder scattering ($\mu_A$: red trace), and the total mobility [$\mu_{TOT}$=($\mu_B^{-1}$+$\mu_R^{-1}$+$\mu_A^{-1}$)$^{-1}$: black trace]\cite{t}. For the fits, we excluded the anomalous outlying point near 4.0$\times$10$^{11}$~cm$^{-2}$.}
	\label{fig1}
\end{figure}

Before discussing the non-linear transport measurements we describe standard characteristics of the 2DEG. As an alternative approach to changing the electron sheet density (n) by applying a voltage to a front or back gate, we could controllably increment n in small steps from $\sim$0.5 to $\sim$4.0$\times$10$^{11}$~cm$^{-2}$ by careful illumination of the Hall bar at $\sim$0.3~K with a standard red light emitting diode passing a small current up to a few tens of nano-amperes. Figure 1 shows the 0.3~K values of n and (transport) mobility $\mu$ obtained using a standard lock-in technique with AC excitation current of 100~nA at 13.5~Hz. We note that n here is the Hall carrier density. Data points were collected on three separate cooldowns. The dependence of mobility on electron density is reproducible from cooldown to cooldown except near the highest density $\sim$4.0$\times$10$^{11}$~cm$^{-2}$ attained for maximum illumination that leads to saturation of the persistent photoconductivity effect~\cite{KANE1986}. We note that n=4.0$\times$10$^{11}$~cm$^{-2}$ is well below the electron density required to populate the first excited sub-band of the QW (estimated to be 1.8$\times$10$^{12}$~cm$^{-2}$ based on a simple calculation using approrpriate material parameters). Empirically we find that the observed sub-linear dependence of mobility on electron density follows a power-law relationship $\mu_{emp}{\sim}$(n-n$_0$)$^{0.42}$ where n$_0$ is a constant: see magenta trace in Fig.~1. The observed dependence is qualitatively similar to those reported in earlier works on InGaAs QWs (see for example Refs.~\onlinecite{KANE1986, GOLD1988, RAMVALL1998}). The dependence is consistent with theoretical models for disorder scattering mechanisms in 2DEG structures~\cite{GOLD1988, DASSARMA2014}. At low electron density, $\mu$ is expected to increase with n when limited by ionized impurity scattering. At high electron density, $\mu$ is expected to saturate in quantum wells \cite{brum1985self} (or even decrease with n in heterojunctions \cite{bastard1983energy}), when limited by alloy disorder scattering. Following the model in Ref.~\onlinecite{DASSARMA2014} which includes screening, we have fitted our data incorporating the appropriate scattering mechanisms. Figure~1 includes calculated mobility versus density traces for background ionized impurity scattering ($\mu_B$), remote ionized impurity scattering ($\mu_R$), short range alloy disorder scattering ($\mu_A$), and the total mobility $\mu_{TOT}$=($\mu_B^{-1}$+$\mu_R^{-1}$+$\mu_A^{-1}$)$^{-1}$ \cite{t}.

For the experiments discussed below, we focus on data taken during the third cooldown (\#3) for which n is in the range 1.6-3.9$\times$10$^{11}$~cm$^{-2}$. Over this range, from the determined values of the transport mobility, we find that the transport mean free path (transport lifetime) increases from 1.2~$\mu$m (4.8~ps) to 2.7~$\mu$m (7.1~ps). For the transport lifetime we took the effective mass (m$^*$) to be 0.047m$_0$ where m$_0$ is the free electron mass~\cite{STUD2012}. From other measurements (data not shown~\cite{VICTORTHESIS}) we also determined the quantum mobility $\mu_q$ and subsequently the quantum lifetime $\tau_q$ by Dingle plot analysis of the low B-field Shubnikov-de Haas (SdH) envelope~\cite{STUD2012, COL1991}. $\mu_q$ ($\tau_q$) is found to be $\sim$12000 cm$^2$/Vs ($\sim$0.31~ps) for n in the range 1.6-3.2$\times$10$^{11}$~cm$^{-2}$ but rises to $\sim$19000~cm$^2$/Vs ($\sim$0.51~ps) near 3.9$\times$10$^{11}$~cm$^{-2}$ when the electron density is close to saturating. From $\tau_q$ we can then estimate the Landau level broadening $\Gamma$. Defining the full width at half maximum (FWHM) of a Lorentzian-shaped Landau level to be 2$\Gamma$~\cite{STUD2012}, we find $2\Gamma$=$\hbar$/$\tau_q$ to be in the range $\sim$1.2-2.0~meV. Note for this Hall bar the longitudinal resistance becomes very large at low density n=n$_0{\sim}$0.5$\times$10$^{11}$~cm$^{-2}$ when the 2DEG is almost insulating and the mobility collapses (see Fig.~1). This density (marking the percolation threshold for transport) equates to a Fermi energy of $\sim$2~meV. From this observation, we conclude that there are effective random potential fluctuations of amplitude $\sim$2~meV. 

\section{DIFFERENTIAL RESISTANCE COLORMAPS}

A map (or ``phase diagram") displaying how the non-linear transport properties of quantum Hall breakdown depend on current (I) and B-field can be biult up by sweeping the current and stepping the B-field. Colormaps of directly measured as opposed to numerically derived longitudinal and transverse differential resistances r$_{xx}$ and r$_{xy}$ respectively are shown in Fig.~2 for n=1.6$\times$10$^{11}$~cm$^{-2}$. The four-point measurements of the differential resistances r$_{xx}$(I$_{DC}$)$\equiv$dV$_{xx}$/dI and r$_{xy}$(I$_{DC}$)$\equiv$dV$_{xy}$/dI were performed simultaneously by driving a combination of a DC current I$_{DC}$ and a small AC excitation current of 100~nA at 13.5~Hz through the Hall bar and measuring the AC voltage component drop $\Delta$V$_{xx}$ and $\Delta$V$_{xy}$ between appropriate potential probes of the Hall bar using a standard lock-in technique~\cite{ZHANG2009, STUD2012}.

\begin{figure}[!h]
	\centering
	\includegraphics[width=0.5\textwidth]{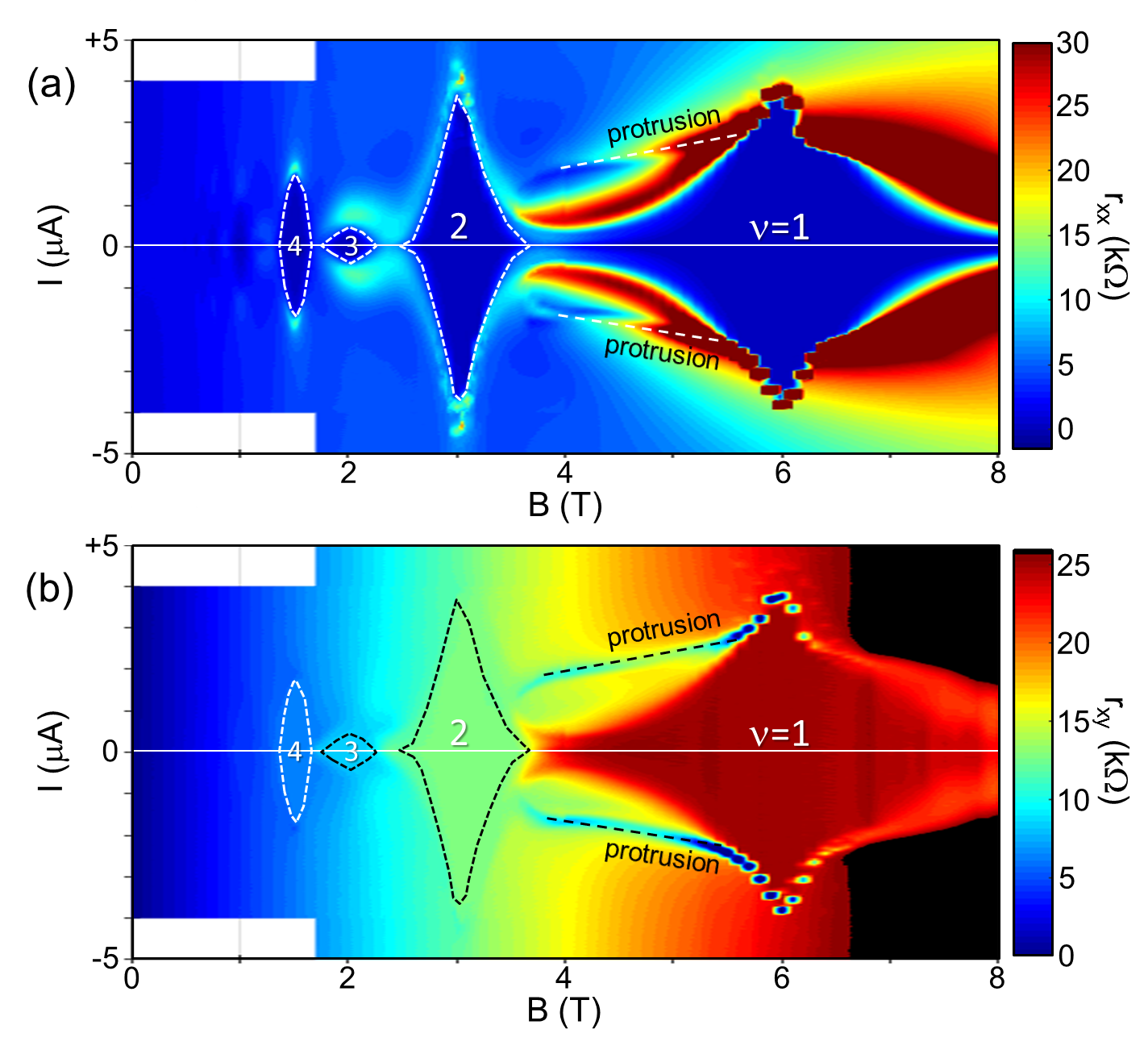}\\
	\caption{Experimental colormaps of (a) r$_{xx}$ and (b) r$_{xy}$: n=1.6$\times$10$^{11}$~cm$^{-2}$, $\mu$=170000~cm$^2$/Vs. Note the regions of r$_{xy}$ exceeding 26~k$\Omega$ are colored black. In both panels the $\nu$=2, 3 and 4 transport diamonds are outlined, and the high-current protrusion features emerging from near the high current tips of the $\nu$=1 transport diamond are identified. The colormaps are built-up with three overlying data sets. The current sweep rate is typically a few tens of nano-amperes per second. The B-field step at low (high) B-field is 25~mT (100~mT).}
	\label{fig2}
\end{figure}

The distinctive dark blue diamond-shaped regions in the map of r$_{xx}$ identify where r$_{xx}{\sim}$0. These regions are the so-called transport diamonds~\cite{STUD2012}. In the map of r$_{xy}$ too, diamonds are visible within which the r$_{xy}$ is quantized. The diamonds in r$_{xx}$ and r$_{xy}$ occupy the same region in the I-B plane. We can now assign these transport diamonds to specific integer filling factors from the value of r$_{xy}$ consistent with the expected quantized values of the Hall resistance. We conclude that the two most prominent diamonds centered at 3~T and 6~T in Fig.~2 are related to filling factors $\nu$=2 and 1 respectively. 

We note the following about the transport diamonds revealed in Fig.~2: (i) a horizontal cut through the diamonds in r$_{xx}$ near I$_{DC}{\sim}$0 would essentially give a traditional plot of R$_{xx}$=V$_{xx}$/I$_{DC}$ with swept B-field when measured in the low current linear response regime (R$_{xx}$=r$_{xx}$) and reveals zeros in R$_{xx}$ at integer filling factors at high B-field in the quantum Hall regime and SdH minima at low B-field; (ii) with increasing B-field, the diamonds become more pronounced and bigger (most clearly reflected in the maximum width in current of the diamonds); (iii) on passing from inside a diamond to outside a diamond, the quantum Hall effect breaks down; and (iv) beyond the diamonds, the quantum Hall effect has broken down and transport is generally non-linear (R$_{xx}{\neq}$r$_{xx}$).

The colormaps provide a wealth of information in the strong quantum Hall regime beyond that reported in Ref.~\onlinecite{STUD2012} which describes transport diamonds for an InGaAs/InP Hall bar but only for B-fields up to 5~T and for n=5.3$\times$10$^{11}$~cm$^{-2}$, so filling factors no lower than $\nu$=5 were investigated. We found it insightful to capture colormaps over a wide range of B-field and for sheet current densities up to $\sim$1~A/m. The colormaps in Fig.~2 together with those presented in the next section for progressively higher electron density provide access to a wide range of filling factor diamonds from $\nu$=1 to at least 10. We note a colormap of V$_{xx}$ showing the $\nu$=2 diamond is presented in Ref.~\onlinecite{PANOS2014} for a GaAs/AlGaAs Hall bar of width 10~$\mu$m, and colormaps of differential resistance between $\nu$=2 and 4 are presented in Refs.~\onlinecite{BAER2015, ROSS2016} for high mobility GaAs/AlGaAs structures of width 500~$\mu$m or greater and for sheet current densities much less than 1~A/m.
	 
In particular, with the capability to measure the $\nu$=1 transport diamond we uncovered distinctive high-current features that emerges from near the high current tips of the diamond. These ``protrusion" are clear in Fig. 2 as peaks in r$_{xx}$ and valleys in r$_{xy}$ that track to lower current as the B-field is reduced, and end near the high B-field side of the $\nu$=2 diamond. Note the value of r$_{xy}$ in the valley is $\sim$h/2e$^2$ or lower. The protrusion features were also seen in a second Hall bar device made from a similar In$_{0.76}$Ga$_{0.24}$As/In$_{0.53}$Ga$_{0.47}$As/InP hetero-structure (data not shown~\cite{VICTORTHESIS}). The protrusions are discussed further in Secs. IV and V.

\section{ELECTRON DENSITY DEPENDENCE OF QUANTUM HALL BREAKDOWN FEATURES}

\begin{figure*}[htbp]
	\centering
	\includegraphics[width=1\textwidth]{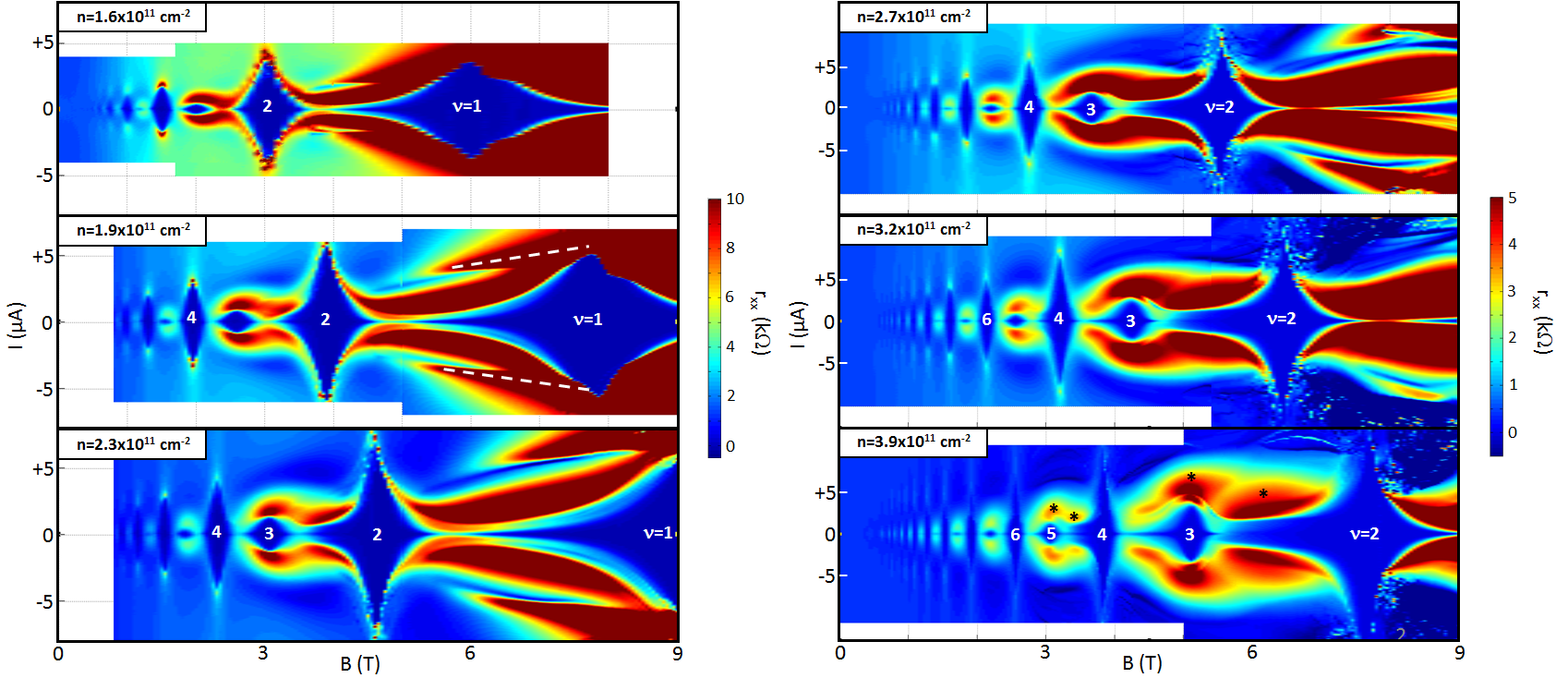}
	\caption{Experimental colormaps of r$_{xx}$ for six electron densities in the range of n=1.6$\times$10$^{11}$~cm$^{-2}$ to n=3.9$\times$10$^{11}$~cm$^{-2}$ attained after controlled illumination. Note for the left- (right-) side panels the color scales and the current axis ranges are different. White dashed lines through the center of protrusions are marked on the colormap for n=1.9$\times$10$^{11}$~cm$^{-2}$. Asterisks identify distinctive ``double-peak" features for positive current polarity in the vicinity of the $\nu$=3 and 5 transport diamonds in the colormap for n=3.9$\times$10$^{11}$~cm$^{-2}$. The colormap for n=3.9$\times$10$^{11}$~cm$^{-2}$ is replotted in Fig.~4 with a different colorscale to highlight certain features more clearly.  }
	\label{fig3}
\end{figure*}


\begin{figure*}[htbp]
	\centering
	\includegraphics[width=1\textwidth]{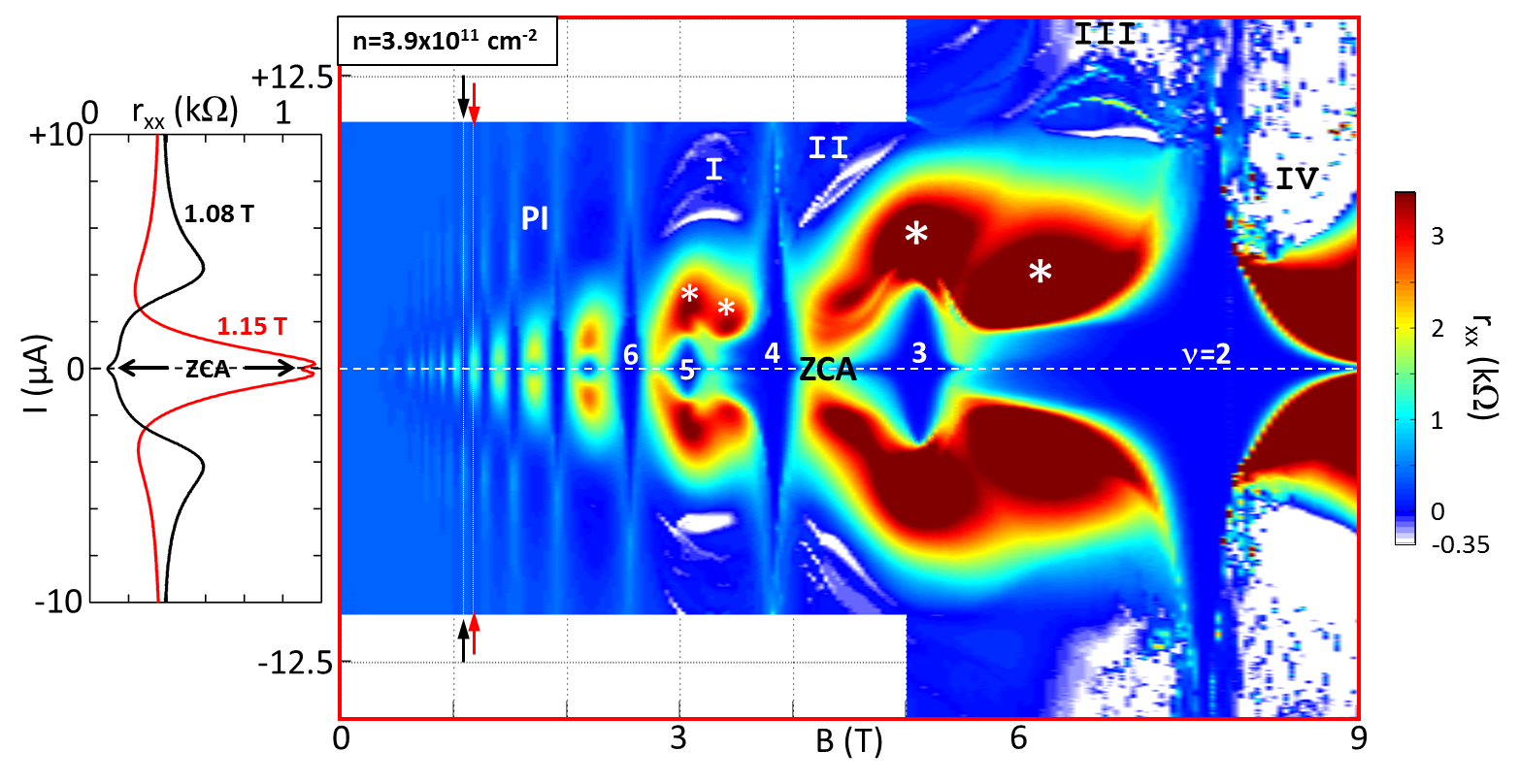}
	\caption{Colormap of r$_{xx}$ for the highest electron density n=3.9$\times$10$^{11}$~cm$^{-2}$ replotted with a different colorscale to emphasize certain features. Phase inversion (PI) of SdH oscillations due to electron heating is observed at low B-field (high filling factor). Left panel illustrates PI effect. The black (red) r$_{xx}$ trace at 1.08~T (1.15~T) cutting through $\nu$=14 SdH minimum (neigboring SdH maximum) at zero current develops into maximum (minimum) at $\sim$4~$\mu$A ($\sim$3~$\mu$A). The zero current anomaly (ZCA) ``dip" discussed further in Sec. VIII is also clear in these two traces. Asterisks in the colormap identify the distinctive ``double-peak" features in the vicinity of the $\nu$=3 and 5 transport diamonds. Regions exhibiting negative differential resistance appear white (r$_{xx}{\leq}$350~$\Omega$) and become widespread beyond 3~T (see, for example, the cresent-shaped features marked I and II between 3 and 5~T. Concurrently, instability appearing as random ``spots" of color also develops above 6~T near the $\nu$=2 diamond (see region marked III) and becomes very pronounced on the high B-field side of the diamond (see region marked IV). }
	\label{fig4}
\end{figure*}

Figure~3 shows measured colormaps of r$_{xx}$ for six electron densities from the lowest (n=1.6$\times$10$^{11}$~cm$^{-2}$) to the highest (n=3.9$\times$10$^{11}$~cm$^{-2}$). Several noteworthy trends are clear with increasing electron density:

\begin{enumerate}[leftmargin=0.5cm]

\item The transport diamonds shift systematically to higher B-field as expected. For example, the center of the $\nu$=2 diamond shifts from $\sim$3~T at n=1.6$\times$10$^{11}$~cm$^{-2}$ to $\sim$7.5~T at n=3.9$\times$10$^{11}$~cm$^{-2}$. Note that the $\nu$=1 diamond is substantially in range only for the first two densities. 

\item The size of the transport diamonds increases. For example, the maximum critical current of the $\nu$=2 diamond increases from $\sim$4~$\mu$A at n=1.6$\times$10$^{11}$~cm$^{-2}$ to $\sim$11~$\mu$A at n=3.9$\times$10$^{11}$~cm$^{-2}$. We will return to the subject of how the critical currents for even- and odd-filling factors vary with B-field and electron density in Sec.~VII.

\item The phenomenon of phase inversion of SdH oscillations at high filling factors discussed in Ref.~\onlinecite{STUD2012} due to electron heating is observed at low B-field and high current, and the phase inversion becomes more visible at higher electron density. By phase inversion we mean minima (maxima) of SdH oscillations in r$_{xx}$ versus B-field at zero current develop into maxima (minima) at high current (see also Fig.~4).
 
\item High even-filling factor ($\nu{\geq}$6) transport diamonds reported in Ref.~\onlinecite{STUD2012} appeared to have straight edges. This would seem to be the case too in Fig.~3 but closer inspection reveals otherwise for low-filling factor diamonds. For example, the $\nu$=4 diamond appears to have straight edges at low density, but these edges become curved at higher density. The edges of the $\nu$=1 and $\nu$=2 diamonds are clearly curved for all densities shown~\cite{u}.

\item Regarding odd-filling factors, at n=1.6$\times$10$^{11}$~cm$^{-2}$, only the $\nu$=1 transport diamond is well developed. At n=3.9$\times$10$^{11}$~cm$^{-2}$, the $\nu$=3 and $\nu$=5 diamonds become clear too. The $\nu$=1 diamond is out of range beyond 9~T.

\item Outside the $\nu$=1 transport diamond, features where r$_{xx}{\geq}$10~k$\Omega$ protrude towards lower B-field are clear for the lower densities.These protrusions always start from near the finite current tips of the $\nu$=1 diamond and appear to track towards zero current at 0~T. They also weaken and eventually terminate at a point located approximately midway along the high B-field sides of the $\nu$=2 diamond (for example, near 4.3~T for n=1.9$\times$10$^{11}$~cm$^{-2}$). We will discuss these protrusions later in Sec. V in connection to Fig.~5.

\item At higher density, the regions just outside the $\nu$=3 and $\nu$=5 transport diamonds develop a distinctive ``double-peak" structure. These double-peak features where r$_{xx}\geq$5~k$\Omega$ are clearest at n=3.9$\times$10$^{11}$~cm$^{-2}$ (in the colormap, the peaks are marked by asterisks on the positive current side). For both the $\nu$=3 and $\nu$=5 diamonds, the leftmost peak is near the finite current tip of the diamond and the rightmost peak is located on the high B-field side of the diamond. The origin of the curious double-peak features is unknown.

\item Starting at n=2.7$\times$10$^{11}$~cm$^{-2}$, close inspection of the colormaps on the right side of Fig.~3 reveals that outside the transport diamonds (when the quantum Hall effect has broken down), regions appear where: i.~r$_{xx}$ approaches zero or even becomes negative, i.e., negative differential resistance is exhibited, and ii.~r$_{xx}$ is unstable. These features become even more pronounced at higher density. For clarity, in Fig.~4 we have replotted the colormap for n=3.9$\times$10$^{11}$~cm$^{-2}$ with a colorscale that enhances these features. Regions appearing white where r$_{xx}{\leq}$0 become widespread beyond 3~T, and instability develops near the $\nu$=2 diamond especially on the high B-field side (see the random ``spots" of color). We will examine these features of electric instability in more detail elsewhere~\cite{YU1}.

\item Although integer filling factor diamonds are clear in Fig.~3, fractional filling factor diamonds are not observed even for n=3.9$\times$10$^{11}$~cm$^{-2}$ when the transport mobility is highest (see Fig.~1). The absense of fractional filling factor diamonds in our data contrasts with those observed between $\nu$=2 and 4 in Refs.~\onlinecite{BAER2015, ROSS2016} for high mobility GaAs/AlGaAs hetero-structures ($\mu{\sim}$2$\times$10$^{7}$~cm$^2$/Vs). Based on a simple Coulomb interaction picture~\cite{DU1983, CHANG1983}, we estimate the energy gap for fractional states to be 0.56~meV (0.74~meV) at 4~T (7~T)~\cite{v}. However, the disorder strength in the InGaAs/InP hetero-structure measured is significant ($\hbar/\tau_q{\sim}$1.2-2.0~meV: see Sec.~II), and is comparable to or exceeds the estimated energy gap for fractional states. This accounts for the absence of fractions. Moreover, alloy disorder scattering in the In$_{0.76}$Ga$_{0.24}$As QW channel is likely even more disruptive to the fractional states~\cite{DENG2014}.

\end{enumerate}

\section{SIMPLE PICTURE OF TRANSPORT DIAMONDS}

\begin{figure*}[htbp]
	\centering
	\includegraphics[width=0.9\textwidth]{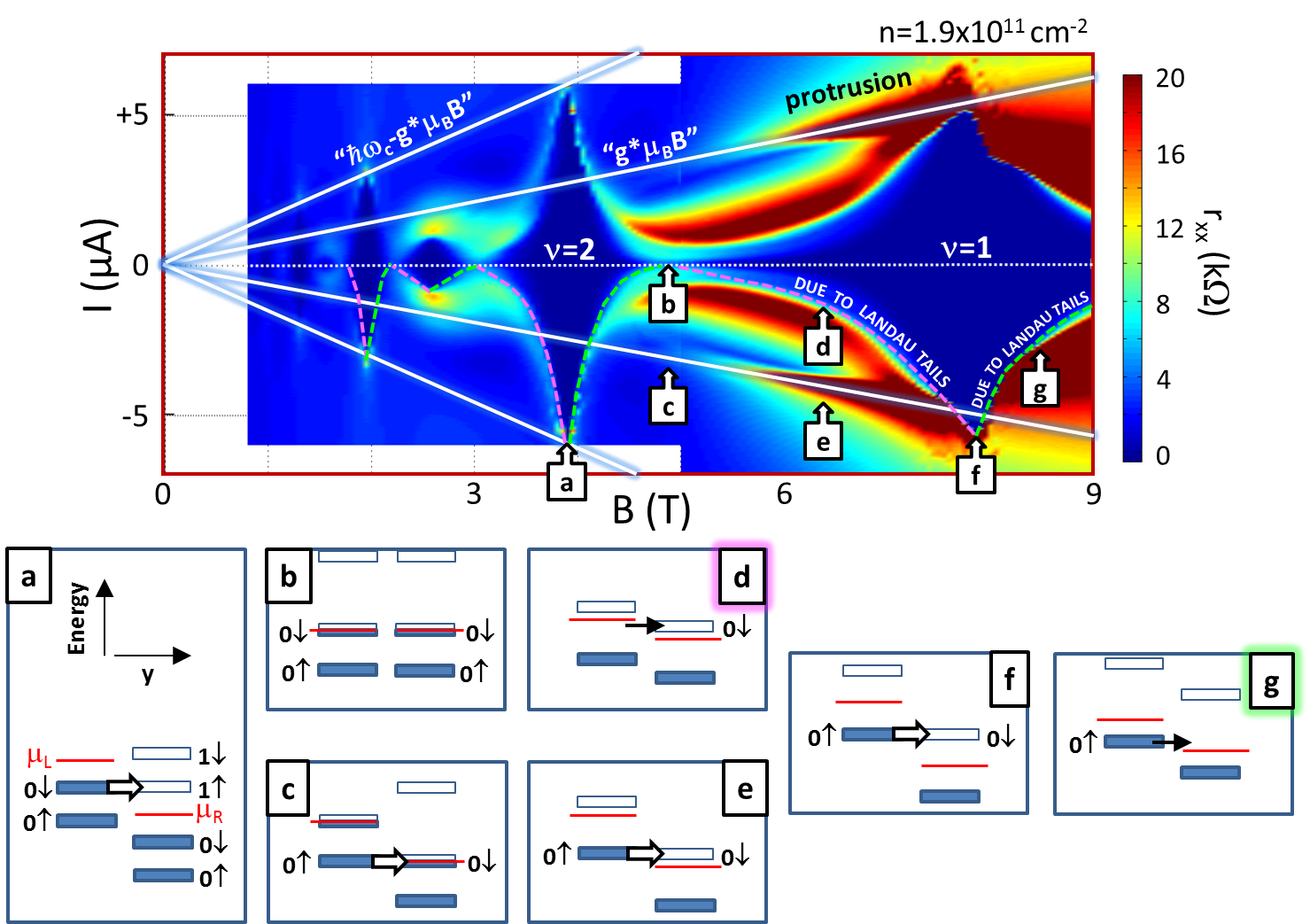}
	\caption{Experimental colormap of r$_{xx}$ for n=1.9$\times$10$^{11}$~cm$^{-2}$. White guide lines bounding tips of even- (odd-)filling factor diamonds are shown and reflect principally the growth in the cyclotron energy (Zeeman energy splitting) with B-field. Cartoons \textbf{a}-\textbf{g} illustrate alignment of Landau levels for the feature marked on the colormap. See text for full description.}
	\label{fig:HB3-simplemodel}
	
\end{figure*}

We now present a simple picture of how we interpret key features related to quantum Hall breakdown that appear most clearly in the colormaps of r$_{xx}$. We will use the colormap of r$_{xx}$ for n=1.9$\times$10$^{11}$~cm$^{-2}$ in Fig.~5 for this purpose. 

It is non-trivial to relate directly current (y-axis in the colormaps) to energy: a point that will be emphasized further in Sec.~VII. However, two features of the transport diamonds stand out. First, the diamond size generally increases with increasing B-field (for odd-filling and even-filling factors separately). Second, the diamond size alternates between relatively small for odd-filling factors and relatively big for even-filling factors. These observations strongly suggest the width of the diamond in current is related to the Zeeman energy splitting ($\mid$g$^*{\mid}\mu_B$B) at odd-filling and to the cyclotron energy ($\hbar\omega_c$=$\hbar$eB/m$^*$) at even-filling. Here, g$^*$, $\mu_B$, $\hbar$, and e respectively are the effective g-factor, Bohr magneton, the reduced Planck constant, and the charge of an electron. For odd-filling factors, we would then expect the maximum width of the (2N+1)th diamond (N= 0, 1, 2, \ldots for $\nu$=1, 3, 5, \ldots) to reflect the transition (N,$\uparrow$)$\rightarrow$(N,$\downarrow$) where the Landau level index N is conserved, but electron spin is flipped (by hyperfine or spin-orbit coupling~\cite{KAWA2011}) from up to down. For even-filling factors, the maximum width of the (2N+2)th diamond (N=0, 1, 2, \ldots for $\nu$=2, 4, 6, \ldots) could reflect the transition (N,$\downarrow$)$\rightarrow$(N+1,$\downarrow$) where N is changed but electron spin is conserved. More likely, it reflects the lower energy transition (N,$\downarrow$)$\rightarrow$(N+1,$\uparrow$) where electron spin is flipped too. This transition has energy $\hbar\omega_c-\mid\mbox{g}^*\mid\mu_B\mbox{B}$. In Fig.~5, we have added white guide lines that separately bound odd- and even-filling factor diamonds and interpolate to zero current at zero field. In Sec.~VII, we will examine more carefully the magnetic field dependence of the critical currents.

The transitions we picture as relevant are principally tunneling transitions between Landau levels that are tilted in the y-direction (across the Hall bar). These transitions are induced by the Hall electric field when current is applied. This Zener-type tunneling mechanism is discussed in Refs.~\onlinecite{EAVES1986, KOMI2000}. Inter-Landau level tunneling is also referred to in the literature as quasi-elastic inter-Landau level scattering~\cite{EAVES1986}. Cartoons \textbf{a}-\textbf{g} in Fig.~5 illustrate alignment of spin-split Landau sub-band energy levels for the features marked on the colormap. In each cartoon, there are two ladders of energy levels. Relevant levels are labeled in compact form, for example, 0$\uparrow$=(0,$\uparrow$). Extended states in each level are shown as bars, and blue (white) identifies filled (empty) states. The left (right) ladder of levels is filled up to the local chemical potential $\mu_L$ ($\mu_R$). L and R represent two points along the y-axis. L and R are defined such that $\mu_L{\geq}\mu_R$ so that transitions occur from the left ladder to the right ladder for either current polarity. In the next section, we will specify further what points ideally L and R represent and discuss the characteristic length scales that could be associated with the separation between these two points.

The specific features marked in Fig.~5 are the following:

\begin{enumerate}[leftmargin=0.5cm]

\item At $\nu$=2, the chemical potential is midway between (0,$\downarrow$) and (1,$\uparrow$). At the tips of the $\nu$=2 diamond [\textbf{a}], breakdown occurs when fully filled (0,$\downarrow$) at L aligns with empty (1,$\uparrow$) at R (cartoon \textbf{a} depicts resonant inter-Landau level transition with spin-flip). Similarly, at $\nu$=1, the chemical potential is midway between (0,$\uparrow$) and (0,$\downarrow$). At the tips of the $\nu$=1 diamond [\textbf{f}], breakdown occurs when fully filled (0,$\uparrow$) at L aligns with empty (0,$\downarrow$) at R (cartoon \textbf{f} depicts resonant transition between spin-split sub-band levels of the N=0 Landau level).

\item \textbf{f} also corresponds to the high B-field end of the protrusion. Cartoons \textbf{e} and \textbf{c} depict alignment for points along the protrusion to lower B-field for the same (0,$\uparrow$)$\rightarrow$(0,$\downarrow$) resonant transition when $\mu_R$ approaches (0,$\downarrow$) [\textbf{e}], and lies through (0,$\downarrow$) [\textbf{c}].

\item At zero current, the $\nu$=1 and $\nu$=2 diamonds touch near 4.9~T (cartoon \textbf{b})~\cite{w}. For small currents, in the vicinity of point \textbf{b}, intra-sub-band transitions occur.

\item Lastly, cartoons \textbf{d} and \textbf{g} respectively represent processes responsible for the edges of the $\nu$=1 diamond on the low (dashed pink line) and high (dashed green line) B-field side. In both cases, localized states in the Landau level sub-band tails are important and the processes occur similarly for the other diamonds. Cartoon \textbf{d} depicts transitions from localized states at L in the high [low] energy tail of (0,$\uparrow$) [(0,$\downarrow$)] to extended states in empty (0,$\downarrow$) at R. Cartoon \textbf{g} depicts transitions from extended states in full (0,$\uparrow$) at L to localized states at R in the high [low] energy tail of (0,$\uparrow$) [(0,$\downarrow$)]: extended states in (0,$\uparrow$) at L are no longer all filled.

\end{enumerate}

We are not aware of any calculation or model that would substantially reproduce, even qualitatively, many of the notable features in the colormap of r$_{xx}$ in Fig.~5 (and colormaps in Fig.~3). We have developed a toy model that does generate a number of general features observed in the experimental colormaps. The model is a one-dimensional tunneling model (see Appendix for further details). The motivation for this approach is that the measured colormap in its entirety primarily reflects the B-field evolution of the Landau levels and the broadening of the Landau levels. This is captured by the model. The model though is not microscopic and it does not explicitly include details of spin-flip mechanisms, i.e., all energetically allowed transitions are permitted whether or not electron spin is flipped. The generated colormaps are also sensitive to the choice of the form of the density-of-states (see Appendix for an example of a colormap generated with our model).

\section{TRANSPORT DIAMONDS IN STRONG QUANTUM HALL REGIME}
\label{diamonds}

Two metrics are frequently quoted in the literature to characterize breakdown: the critical current density J$_c$=I$_c$/w at a given filling factor $\nu$ (where I$_c$ and w respectively are the critical current and Hall bar width), and the corresponding critical Hall electric field E$_c$=hI$_c$/$\nu$e$^2$w (where h is the Planck constant). Although in the following section, we will discuss more carefully how one can define and extract I$_c$, from Fig.~5, we estimate that I$_c{\sim}$6~$\mu$A ($\sim$5~$\mu$A) from the tips of the dark-blue colored $\nu$=2 ($\nu$=1) diamond at ~3.9 T (~7.8 T) for n=1.9$\times$10$^{11}$~cm$^{-2}$. Thus, J$_c$ and E$_c$ are respectively $\sim$0.4~A/m and $\sim$5.2~kV/m ($\sim$0.33~A/m and $\sim$8.6~kV/m) at $\nu$=2 ($\nu$=1). We can compare these numbers with those compiled in Refs.~\onlinecite{KAWAJI1996, JECK2001} for GaAs/AlGaAs 2DEG hetero-structures. Interestingly, the numbers we obtain for the InGaAs/InP QW here are quite similar in value. For example, from Refs.~\onlinecite{KAWAJI1996, JECK2001} we infer E$_c$ is in the range of 2-6~kV/m at 4~T ($\sim$6~kV/m at 8~T) for even-filling (odd-filling) factors. In our estimation of J$_c$ and E$_c$, we took w to be the nominal width of the Hall bar (=15~$\mu$m) and neglected any reduction due to wet etching and side wall depletion.

Assuming the Landau levels are tilted uniformly across the Hall bar, one would anticipate that one could make a crude estimate of the minimum characteristic separation between points L and R ($\delta$ in the toy model described in the Appendix) required to induce the transitions depicted in Fig.~5. We shall examine the estimated value for E$_c$ of $\sim$5.2~kV/m at $\nu$=2 (B=3.9~T) for n=1.9$\times$10$^{11}$~cm$^{-2}$. For the pure Zener tunneling mechanism represented in the Fig.~5 cartoon \textbf{a}, we might expect E$_c{\sim}\Delta$/e$\delta$ where $\Delta$=$\hbar\omega_c$-$\mid$g$^*{\mid}\mu_B$B is evaluated at 3.9~T with appropriate values of 0.047m$_0$ and 4.8 respectively for m$^*$ and $\mid$g*$\mid$~\cite{gfactor}. Hence, a value for the characteristic length scale $\delta{\sim}\Delta$/eE$_c{\sim}$(8.5~mV)/(5.2~kV/m) of 1.6~$\mu$m is determined. Reasonably, $\delta{\ll}$w=15~$\mu$m: we would not expect direct tunneling from one side of the Hall bar to the other side. However, $\delta$ is two orders of magnitude larger than the distance over which we would expect direct inter-Landau level tunneling to be significant. As discussed in Ref.~\onlinecite{KAWAJI1996}, the tunneling distance can be related to the spatial extent of wavefunctions of electrons in the Landau levels. This is comparable to the magnetic length $\ell_B$=25.6~nm/$\sqrt{B}$ with B in units of Tesla. For B=3.9~T, $\ell_B{\sim}$13~nm$\ll \delta$. On the one hand, the values for E$_c$ we determine experimentally are similar to those reported in the literature for GaAs/AlGaAs 2DEGs and are up to two orders of magnitude smaller than expected based on a simple pure Zener tunneling picture (see also Ref.~\onlinecite{KOMI2000}). On the other hand, the only other characteristic length scale we have encountered with a value on the micron-scale for the Hall bar studied is the transport mean free path (see Sec.~II).

While a fully satisfactory answer to the long standing problem of accounting for the empirical values of E$_c$ remains to be found \cite{KAWAJI1996}, we make the following comments. The experimentally derived critical Hall electric fields, equally the critical current densities, should be regarded as \textit{average} quantities across the width of the Hall bar. Furthermore, the Landau levels are not tilted uniformly across the Hall bar. Two effects locally can significantly enhance the electric field. Firstly, the Landau levels bend up strongly near the edges of the Hall bar \cite{HAUG1993, WEIS2011}. Secondly, in reality, there are random potential fluctuations in the 2DEG due to disorder \cite{EAVES2001B, SANUKI2001, HAYA2013}. Lastly, although Zener tunneling is undoubtedly involved in quantum Hall breakdown, it is not the only factor. In the electron heating model of Komiyama and Kawaguchi \cite{KOMI2000}, the role of Zener tunneling is primarily to kick-start an avalanche multiplication of excited carriers the conditions for which depend up on the details of energy balance between gain and loss processes.

To finish this section, we describe how the $\nu$=1 and $\nu$=2 transport diamonds in Fig.~5 can be used to make a crude estimate of the g-factor for $\nu$=1. n=1.9$\times$10$^{11}$~cm$^{-2}$ is the highest electron density for which the $\nu$=1 diamond is substantially visible with the 9 T magnet available in the experiments. Note the energy gap associated with the $\nu$=1 diamond is too large for the g-factor to be determined by performing thermal activation measurements over the accessible temperature range of 270~mK to 1.5~K. Instead, we start with the observation that the maximum breakdown current for the $\nu$=2 diamond is $\sim$6~$\mu$A at 3.9~T. We take this current to be proportional to energy $\hbar\omega_c$-$\mid$g$_e^*{\mid}\mu_B$B$\sim$8.5~meV. The subscript e is to emphasize that the g-factor should be that appropriate for an even-filling factor (we have used the value of 4.8 estimated from separate measurements~\cite{gfactor}). For the $\nu$=1 diamond, the maximum breakdown current is $\sim$5~$\mu$A at 7.8~T. We take this current to be proportional to energy $\mid$g$_o^*{\mid}\mu_B$B. The subscript o is to emphasize that the g-factor should be that appropriate for an odd-filling factor subject to enhancement by many-body electron-electron exchange interactions \cite{STUD2012, ANDO1982, NICHOLAS1988, LEAD1998, KRISH2015}. Assuming the constant of proportionality for converting current to energy is the same for both diamonds, the 5~$\mu$A breakdown current for $\nu$=1 is equated to the energy $\sim$7~meV implying $\mid$g$_o^*{\mid}{\sim}$16. Our estimation of $\mid$g$_o^*{\mid}$ here has neglected Landau level broadening and assumed the breakdown currents quoted above for the $\nu$=1 and 2 diamonds reflect the corresponding resonant transitions (effectively method B to be described in Sec. VII). We will discuss these assumptions further in the following section. Nonetheless, recent calculations of exchange enhancement of g-factors in strained InGaAs/InP hetero-structures suggest values of magnitude exceeding 10 are possible for $\nu$=1~\cite{KRISH2015}. 

\section{B-FIELD DEPENDENCE OF CRITICAL BREAKDOWN CURRENT}

The maximum critical current I$_c$ at integer filling (where the transport diamond is widest), and how it depends on the magnetic field with varying electron density has been of considerable attention for a number of decades. Not only has this subject been of interest for metrological reasons, but in principal the dependence can shed light on the physical mechanisms that drive quantum Hall breakdown: see for example Refs.~\onlinecite{NACHT1999, KOMI2000, KAWAJI1996, JECK2001, MAKA2002}. Controlled illumination allows us to compile sufficient data for different values of n to examine the magnetic field dependence of I$_c$ for even- and odd-filling factors for an InGaAs/InP QW 2DEG: a system other than GaAs/AlGaAs which is usually investigated.

The choice of how one defines the onset of breakdown can influence the value of I$_c$. The following two methods are reasonable and have been used in the literature. For method A, I$_c$ is taken to be the value at which r$_{xx}$ exceeds a small threshold. The value of the threshold though is influenced to a degree by the noise level in the measured r$_{xx}$. Furthermore, deviation from r$_{xx}$=0 above a small threshold may occur for a gentle pre-breakdown increase or for multiple-step transitions prior to abrupt breakdown \cite{EBERT1983, Q}. For method B, I$_c$ is simply given by the position of the r$_{xx}$ maximum usually clear at the high current tip of a transport diamond. Note that this corresponds to where the rise in V$_{xx}$ with current is steepest. This is the most direct method to apply except if r$_{xx}$ exhibits multiple-peak features or becomes unstable which is sometimes the case, as noted in Sec. IV, for high electron density and high B-field (low-filling factor)~\cite{Q}. In general, the value of I$_c$ determined by method A is smaller than that determined by method B however these methods can give similar values if breakdown is very abrupt.

Figure 6 shows the critical current I$_c$ determined by method A [panel (a)] and method B [panel (b)] for even (E)- and odd (O)-integer filling factors for six electron densities in the range 1.6-3.9$\times$10$^{11}$~cm$^{-2}$~\cite{Q}. For method A, since the noise level in r$_{xx}$ near zero current for the largest transport diamonds is a few ohms, we have set the threshold value to be 10~$\Omega$. From Fig.~6(a) for method A, we find that data points essentially fall on distinct common straight lines for even- and odd-filling factors. The common line for even-filling factors is steeper than that for odd-filling factors. The lines clearly do not extrapolate to the origin. The even- (odd-) filling factor line intercepts zero current at $\sim$1.5~T ($\sim$2.5~T). From Fig.~6(b), for method B, although the even- and odd-filling factor families of lines are clear, there is considerably more variation and scatter than for method A, i.e., data points for different electron densities do not fall on a single common line. Even-filling factor points above $\sim$1.5~T for each specific electron density appear to have approximately linear trends that would extrapolate to the origin, although there are deviations from the linear dependence near 1~T. Also, with increasing electron density, I$_c$ for even-filling factors clearly changes more rapidly with B-field. Although less pronounced, a similar behavior is observed too for the odd-filling factor points. The odd-filling factor trends intercept zero current near 1~T.

\begin{figure*}[htbp]
  \centering
    \includegraphics[width=0.9\textwidth]{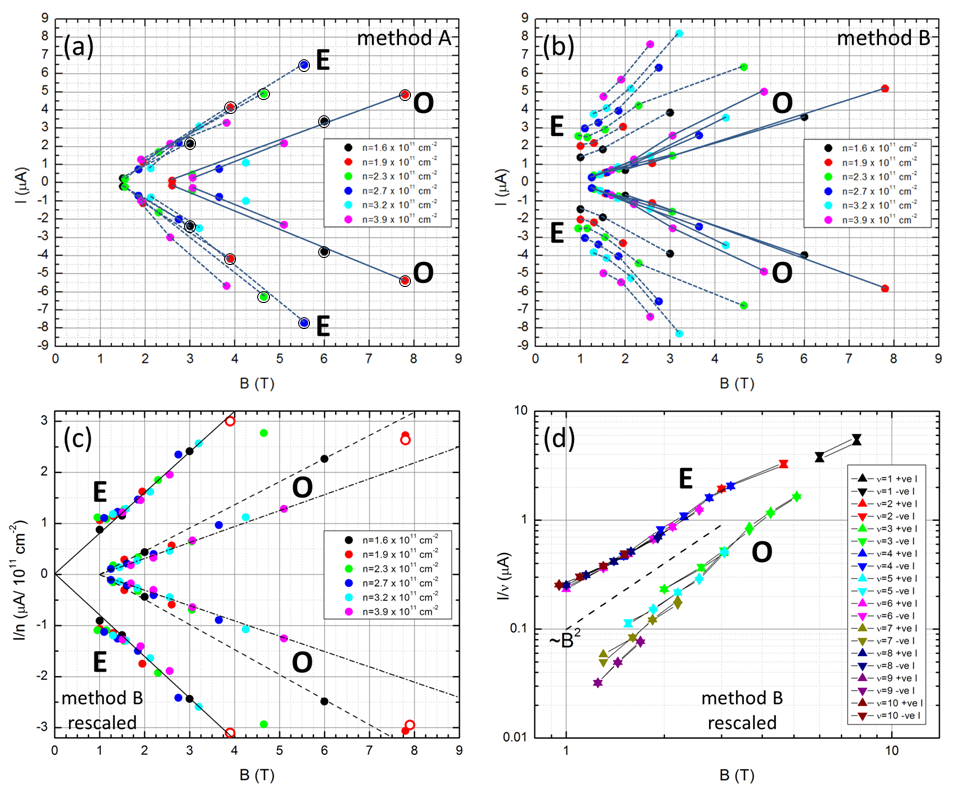}
  \caption{Critical current I$_c$ compiled for even (E)- and odd (O)-integer filling factors for electron densities in the range 1.6-3.9$\times$10$^{11}$~cm$^{-2}$. The values are determined by two methods on examination of r$_{xx}$(I$_{DC}$) traces that are used to build-up the colormaps in Fig.~3. In (a) I$_c$ is the current at which r$_{xx}$ exceeds 10~$\Omega$ (method A), and in (b) I$_c$ is the current position of the r$_{xx}$ maximum at the high current tip of a transport diamond (method B). Guide lines are shown joining two or more data points for even- (dashed) or odd- (solid) integer filling factors at a given electron density. In (a) data points for $\nu$=1 and $\nu$=2 are circled. In (c) panel (b) data points are replotted to show I$_c$/n instead of I$_c$ (method B rescaled). Even-integer filling data points effectively collapse on to a single (solid) line. Odd-integer filling data points lie in a region bounded by the dashed line (with steeper slope for n=1.6$\times$10$^{11}$~cm$^{-2}$ data points) and the dashed-dotted line (with shallower slope for n 3.9$\times$10$^{11}$~cm$^{-2}$ data points): see main text for further discussion. Although not strictly determined by method B, values of I$_c$ estimated from the tips of the dark-blue colored $\nu$=1 and $\nu$=2 diamonds for n=1.9$\times$10$^{11}$~cm$^{-2}$ in Fig.~5 (see Sec. VI) and \textit{rescaled} are also included (open red circles). In (d) panel (b) data points are replotted, on a log-log scale and sorted by $\nu$ instead of n, to show absolute values of I$_c$/$\nu$ (also effectively method B rescaled since $\nu$=nh/eB). The dashed black line with a $\sim$B$^2$ dependence is included as a guide to the eye. In the legend +ve I (-ve I) means positve (negative) current polarity.}
  \label{Fig6}
\end{figure*}

\begin{figure}[htbp]
	\centering
	\includegraphics[width=0.45\textwidth]{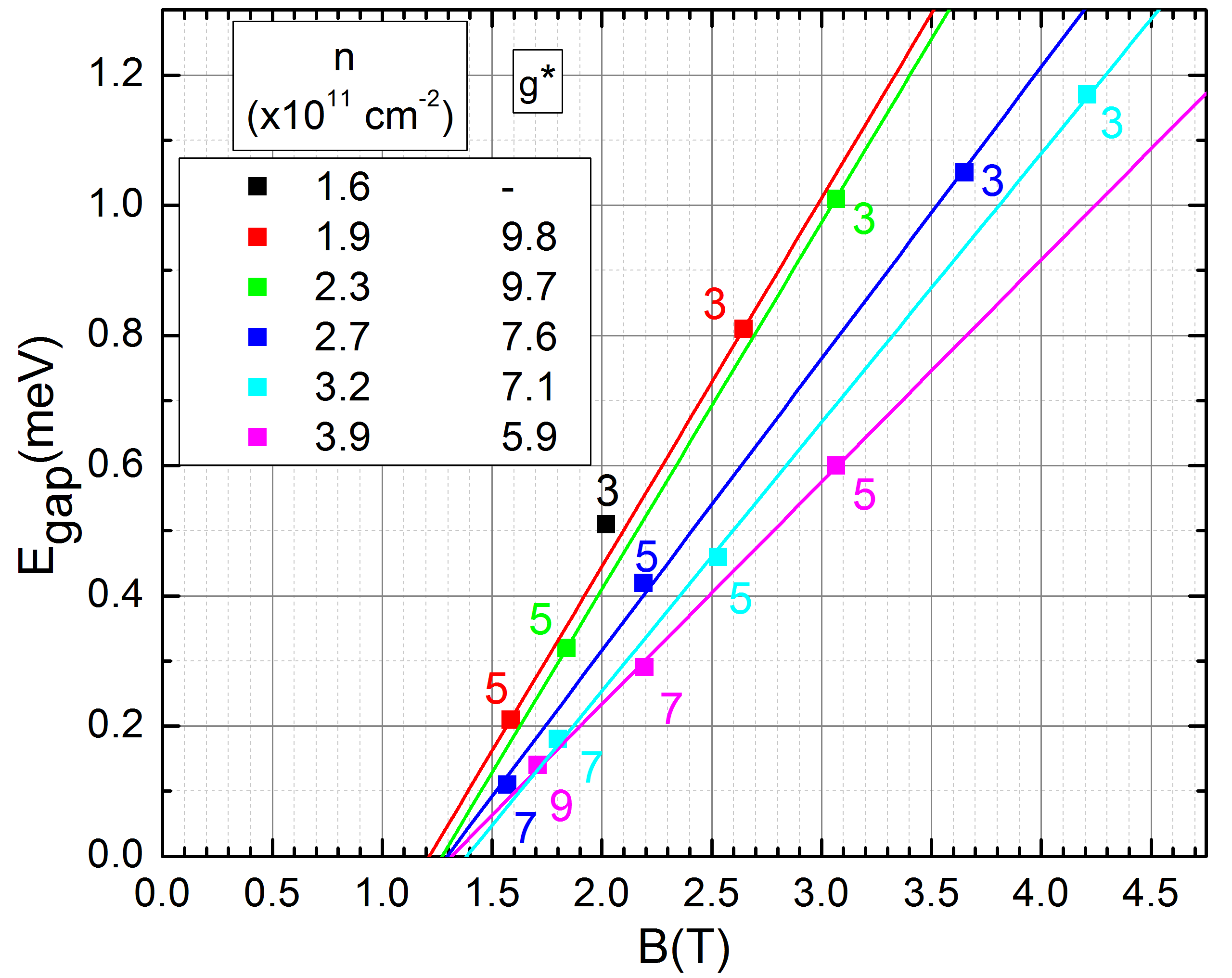}
	\caption{Values of E$_{gap}$ determined from the temperature dependence of r$_{xx}$ at the $\nu$=3, 5, 7, and 9 minima (I$_{DC}$=0) for different electron densities n as a function of B. The number next to a point identifies $\nu$. The lines intercept the B-axis in the range 1.2-1.4~T. Estimated values of $\mid$g$^*{\mid}$ from the slopes are also tabulated. There is no value shown for n=1.6$\times$10$^{11}$~cm$^{-2}$ since only one data point (for $\nu$=3) could be extracted.}
	\label{Fig7}
\end{figure}

We make the following comments. Firstly, the linear dependence with B-field for I$_c$ data points determined by method A in Fig.~6(a) is to a certain degree unexpected. For GaAs/AlGaAs 2DEG hetero-structures it is often argued that the critical Hall electric field E$_c{\propto}$ I$_c$/$\nu$ is proportional to B$^{3/2}$ (for example, see the discussion and plots of compiled data in Refs.~\onlinecite{KAWAJI1996, JECK2001} mostly extracted by method A). However, models for breakdown generally predict values for E$_c$ that are typically at least an order of magnitude larger than experimental values, and depend on factors such as the inclusion of higher order tunneling processes, details of scattering (particularly the B-field dependence of the scattering rate), and Landau level broadening induced by the Hall electric field \cite{KOMI2000, KAWAJI1996}. Others have also suggested a linear B dependence fit of compiled data appears to be at least as good as a B$^{3/2}$ dependence fit \cite{MAKA2002}. For our data, we do not find a power law dependence for even- or odd-filling factors if we examine I$_c$/$\nu$ (data not shown). Consistent with our observations, we note a linear dependence of the $\nu$=1 and 2 breakdown currents with B-field was also recently reported in Ref.~\onlinecite{ALEXANDAR2012} for InSb/AlInSb hetero-structures with mobility exceeding $\sim$160000~cm$^2$/Vs. To emphasize this point, data points for $\nu$=1 and $\nu$=2 are circled in Fig.~6(a). Secondly, the characteristics of I$_c$ determined by method A [Fig.~6(a)] are notably different from the characteristics of I$_c$ determined by method B [Fig.~6(b)] especially regarding the scatter of data points, i.e., method A data points appear to collapse onto two common lines but clearly method B data points do not. However, we stress the two methods reflect two quite different features of Landau levels. Consistent with the discussion in Sec. V, we expect method A to be sensitive to how electrons tunnel between states in the tails of Landau levels (when the tunnel barrier for inter-Landau level transitions becomes sufficiently transparent), and method B to be sensitive to how electrons tunnel resonantly between Landau levels at higher Hall electric field. Since in the framework of our simple picture for the transport diamonds method B offers a means to extract g-factors for odd-filling (crudely done for $\nu$=1 in Sec. VI at n=1.9$\times$10$^{11}$~cm$^{-2}$), we now examine this method more closely after introducing a simple procedure to account for the data scatter. 

At first sight, the scatter in data points in Fig.~6(b) is counter to our intuition. In particular, based on our simple picture described in Sec. V, we take the individual I$_c$ traces for even-integer filling in Fig.~6(b) to reflect the energy $\hbar\omega_c$-$\mid$g$_e^*{\mid}\mu_B$B (see also Sec. VI). This energy is dominated by the cyclotron energy, and is not expected to have any \textit{significant} n-dependence. However, this expectation neglects electrostatic details that determine the proportionality factor between current and energy. Since the Hall electric field is the dominant field, and is proportional to I/n at fixed magnetic field, this prompted us to examine I$_c$/n instead of I$_c$
(method B rescaled): see Fig.~6(c) in which \textit{all} panel (b) data points are replotted. Noticeable immediately, even-integer filling data points now effectively collapse on to a single (solid) line. This universal behavior means I$_c$ is proportional not just to B but also n, i.e., I$_c$/n and not I$_c$ should be equated with energy $\hbar\omega_c$$-$$\mid$g$_e^*{\mid}\mu_B$B. Regarding the odd-integer filling data points, rescaling reduces the scatter to a degree and they lie in a region bounded by the dashed line (with steeper slope) and the dashed-dotted line (with shallower slope)~\cite{SCATTER}. Interestingly, the order by slope of the rescaled odd-integer filling data point traces is inverted with respect to those prior to rescaling [see Fig.~6(b)]. The linear B-dependence trace now appears steepest (shallowest) for n=1.6$\times$10$^{11}$~cm$^{-2}$ (n=3.9$\times$10$^{11}$~cm$^{-2}$), i.e., there is a trend for the slope to decrease with increasing n. Values of I$_c$ estimated from the tips of the dark-blue colored $\nu$=1 and $\nu$=2 diamonds for n=1.9$\times$10$^{11}$~cm$^{-2}$ in Fig.~5 (see Sec. VI) and \textit{rescaled} are also included in Fig 6 (c): see open red circles. Although not strictly determined by method B, since breakdown is fairly abrupt, these extra points nonetheless lie either very close to data points extracted by method B at the corresponding density in the case of $\nu$=1, or very close to the single common line for even-filling factor points in the case of $\nu$=2. 

The rescaling procedure also implies the critical Hall field E$_c$ is proportional to B$^2$ when method B is applied since empirically I$_c{\propto}$Bn (E$_c$=hI$_c$/$\nu$e$^2$w=BI$_c$/new). To corroborate this, in Fig.~6(d) \textit{all} panel (b) data points are replotted, on a log-log scale and sorted by $\nu$ instead of n, to show absolute values of I$_c$/$\nu$ (also effectively method B rescaled). The dependence we see for both even- and odd-filling is closer to B$^2$ than B or B$^{3/2}$. Interestingly, this is similar to observations for Hall induced resistance oscillations (HIROs) in high mobility GaAs/AlGaAs 2DEGs at low magnetic fields ($<$0.5~T), where Zener tunneling between Landau levels takes place across distance 2R$_c\sim$1/B (the cyclotron diameter) \cite{yang2002zener}. If we equate the characteristic length scale $\delta$ for Zener tunneling introduced in Sec. VI to $2R_c$, this would lead to a B$^2$ dependence for E$_c$. However, in the strong quantum Hall regime at large magnetic fields, 2R$_c$ is much smaller than the length of 1.6 $\mu$m at 3.9~T that we estimated earlier. We stress the rescaling in Fig.~6(c) and Fig.~6(d) involves no fitting or knowledge of quantities other than n or $\nu$.

Having demonstrated that data points of I$_c$/n for even-integer filling collapse on to a single line which can be equated to the energy $\hbar\omega_c$ -$\mid$g$_e^*{\mid}\mu_B$B, we proceed to estimate g-factors for odd-filling from Fig.~6(c). We now expect the individual I$_c$/n traces for odd-integer filling to be a measure of the energy $\mid$g$_o^*{\mid}\mu_B$B (see also Secs. V and VI). By examining the ratio of the slopes of the single solid line for even-filling and the dashed and dashed-dotted bound lines for odd-filling we estimate $\mid$g$_o^*{\mid}$ to be in the range $\sim$15-21 if we take values of 0.047m$_0$ and 4.8 respectively for m$^*$ and $\mid$g$_e^*{\mid}$. The value for the g-factor appears to be larger (smaller) for lower (higher) density~\cite{AAA}. In contrast to $\mid$g$_e^*{\mid}$, not only is a larger value for $\mid$g$_o^*{\mid}$expected but an n-dependency is reasonable too. The increase of $\mid$g$_o^*{\mid}$ with decreasing n is significant and likely reflects an increase in the strength of the effective electron-electron interactions \cite{ANDO1982, LEAD1998, SADOFYEV2002}. To investigate this point further, we end this section by comparing the effective exchange-enhanced g-factors estimated for odd-filling from the rescaled compiled data in Fig.~6(c) to those determined by conventional thermal activation measurements over the same range of electron density.

At exact odd-integer filling, the Fermi level is midway between two broadened spin-split Landau sub-bands (N,$\uparrow$) and (N,$\downarrow$). Close to equilibrium (I$_{DC}$=0 and small AC excitation current), the energy to activate electrons from the Fermi level to extended states in the broadened level (N,$\downarrow$) is E$_{gap}$/2, where E$_{gap}$=$\mid$g$_o^*{\mid}\mu_B$B-2$\Gamma$ [the effective energy gap between (N,$\uparrow$) and (N,$\downarrow$) \cite{STUD2012, KRISH2015, ANDO1982, NICHOLAS1988, LEAD1998, USHER1990, ZHANG2006, LI2016}. This activation energy can be extracted from the temperature (T) dependence of r$_{xx}$ at odd-filling factor minima, namely from an Arrhenius plot if $\textrm{r}_{xx}{\sim}\textrm{exp(-E}_{gap}/2\textrm{k}_B\textrm{T})$ where k$_B$ is the Boltzmann constant. We examined the temperature dependence of r$_{xx}$ at I$_{DC}$=0 for odd-filling factor minima as a function of B. Plots of ln(r$_{xx}$) vs. 1/T for T in the range 0.3~K to $\sim$1.6~K were found to be close to linear for T$\geq$1~K for electron densities in the range 1.6-3.9$\times$10$^{11}$~cm$^{-2}$ (data not shown \cite{VICTORTHESIS}). Fitting the plots at higher temperature where they are reasonably linear, E$_{gap}$ was extracted. The values  of E$_{gap}$ are plotted in Fig.~7 as a function of B (B-field at which the r$_{xx}$ minima are located). From this plot, except for the lowest density, $\mid$g$_o^*{\mid}$ can be estimated from the linear dependence on B \cite{ASSUMP}, while $\Gamma$ can be estimated from the zero B-field intercept \cite{LEAD1998, USHER1990, DOLGOPOLOV1997, ZHANG2006, LI2016}. For n=1.9, 2.3, 2.7, 3.2, and 3.9$\times$10$^{11}$~cm$^{-2}$, $\mid$g$_o^*{\mid}$ is determined to be 9.8, 9.7, 7.6, 7.1, and 5.9 respectively. The uncertainty in these numbers is estimated to be $\pm$0.4. This number reflects uncertainty in the slopes of the fit lines for n=2.7, 3.2, and 3.9$\times$10$^{11}$~cm$^{-2}$ for which there are three data points in each case. 

The trend of increasing $\mid$g$_o^*{\mid}$ with decreasing n suggested from the analysis of the rescaled method B data in Fig.~6(c) is also seen in the thermal activation data in Fig.~7. The values of the exchange-enhanced g-factor determined by thermal activation measurements here are consistent with those extracted by the same technique and reported in earlier work for similar materials \cite{STUD2012, KRISH2015}. Notable, however, the estimated values for $\mid$g$_o^*{\mid}$ from the rescaled method B approach are a factor of $\sim$2-3 times higher than those from the thermal activation measurements. Currently we do not have a full detailed understanding of the difference. However, the exchange energy depends on the difference in population of spin-up and spin-down sub-bands \cite{ANDO1982, NICHOLAS1988, LEAD1998}. This difference in population changes with temperature and so the effective g-factor determined from a thermal activation measurement is expected to be smaller than the effective g-factor determined from the transport diamonds at fixed base temperature. Lastly, from the zero B-field intercepts in Fig.~7, we find $2\Gamma$ to be in the range $\sim$0.4-0.8~meV. These numbers are about a factor of 2 lower than those given in Sec. II that were determined by the more well established Dingle plot method. Our estimation of $\Gamma$ from analysis of Fig.~7 though neglects any contribution to conduction by mechanisms such as variable range hopping \cite{ANDO1982, BUSS2005}, and assumes the boundary in energy between localized and extended states in a density-of-states peak is well defined.

\section{ZERO CURRENT ANOMALY}

\begin{figure}[htbp]
	\centering
	\includegraphics[width=0.5\textwidth]{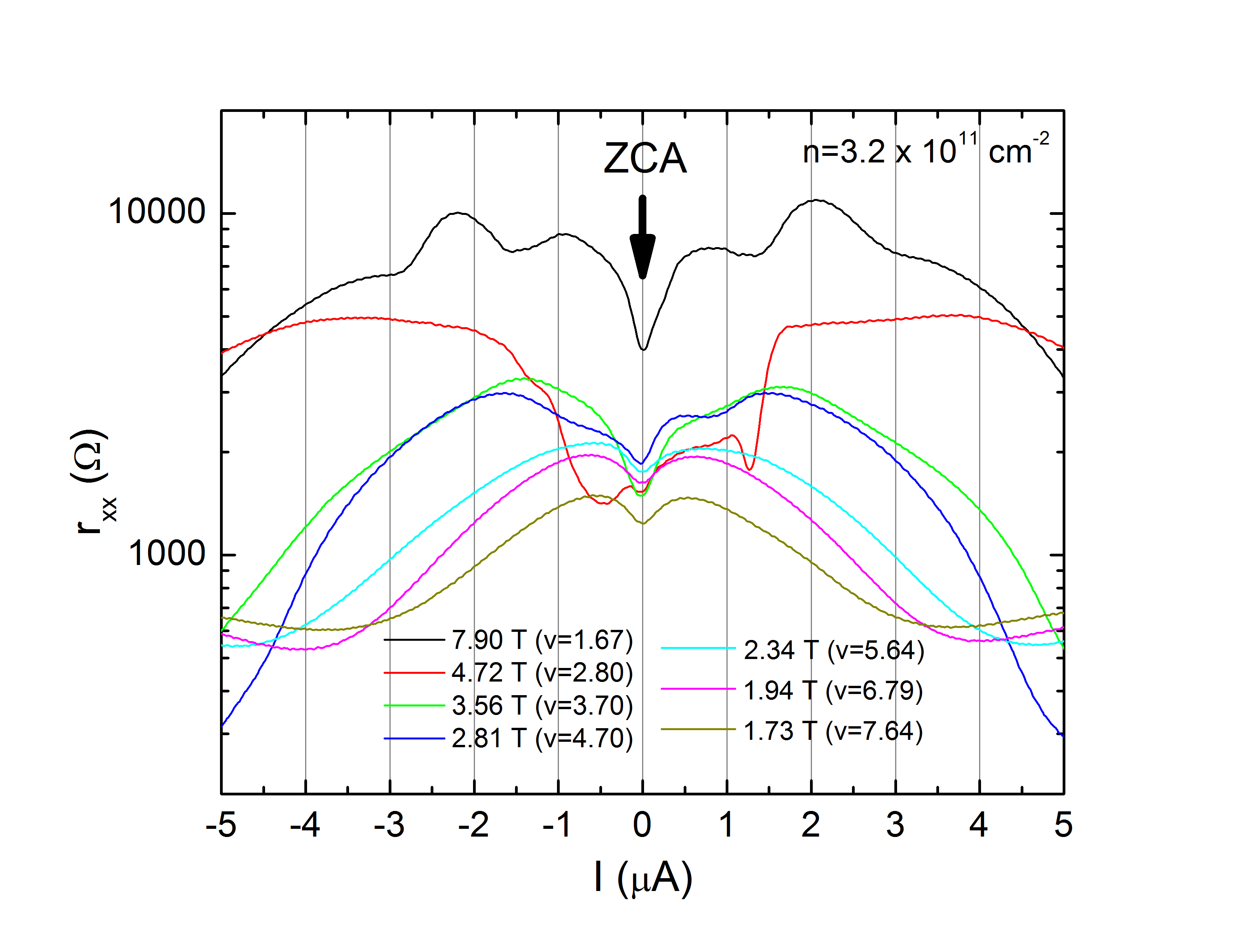}
	\caption{r$_{xx}$ sections through colormap for B-fields near where neighboring transport diamonds touch for electron density n=3.2$\times$10$^{11}$~cm$^{-2}$. See r$_{xx}$ colormap in Fig.~3. The zero current anomaly (ZCA) ``dip" is labeled and is present in all traces shown.}
	\label{fig:ZCA}
\end{figure}

	Close inspection of the r$_{xx}$ colormaps in Fig.~3 reveals a narrow ``dip" in r$_{xx}$ in regions close to zero current most notably where the transport diamonds touch. This dip is a signature of the zero current anomaly (ZCA) described in Ref.~\onlinecite{STUD2012}. In fact, the ZCA can be observed over a wide range of B-field provided that r$_{xx}$ is not zero in the vicinity of zero current over a current range exceeding $\sim$0.4~$\mu$A (see also Fig. ~4). To illustrate this, Fig.~8 shows r$_{xx}$ sections on a log scale for n=3.2$\times$10$^{11}$~cm$^{-2}$ that cut through the points close to where neighboring transport diamonds touch at zero current. At such points, encompassing the $\nu$=1 to 8 transport diamonds, the ZCA is clear and has FWHM of typically $\sim$0.4~$\mu$A~\cite{y}.
	
	Studenikin~\textit{et al.}~\cite{STUD2012} described basic properties of the ZCA for InGaAs/InP Hall bars, but only for B-fields up to $\sim$5~T and for an electron density $\sim$5.3$\times$10$^{11}$~cm$^{-2}$, so filling factors no lower than $\nu$=5 were investigated. The ZCA FWHM was found to be $\sim$2.1~$\mu$A ($\sim$0.25~$\mu$A) for a 100~$\mu$m (10~$\mu$m) wide Hall bar, so this indicates, consistent with the FWHM ~0.4~$\mu$A for the 15~$\mu$m wide Hall bar measured here, that the FWHM is proportional to the Hall bar width~\cite{z}. From the near linear increase in width of even-filling factor transport diamonds with B-field which principally reflects the growth in cyclotron energy (see discussion in Secs. V-VII), we can convert the FWHM of the ZCA into an energy set by $\hbar\omega_c$-$\mid$g$^*{\mid}\mu_B$B. For an effective mass of 0.047m$_0$ and taking $\mid$g$_e^*{\mid}{\sim}$4.8 we estimate that the FWHM $\sim$0.4~$\mu$A here is equivalent to an energy $\sim$0.3~meV.
	
	Reference~\onlinecite{STUD2012} also reported the ZCA in higher mobility GaAs/AlGaAs Hall bar structures so the phenomenon appears to reflect intrinsic properties of a 2DEG. The origin of the ZCA is not understood and we are not aware of any theoretical description on this subject. In Ref.~\onlinecite{STUD2012}, it is speculated to originate from modifications to the density-of-states of interacting electrons in the presence of disorder and a magnetic field.

\section{Conclusions}

We have investigated non-linear magneto-transport in a Hall bar device made from a strained In$_{0.76}$Ga$_{0.24}$As/In$_{0.53}$Ga$_{0.47}$As/InP quantum well hetero-structure (with mobility approaching 300000~cm$^2$/Vs for n$\sim$4.0$\times$10$^{11}$~cm$^{-2}$ at 0.3~K). We have demonstrated that maps of the differential resistance (r$_{xx}$ and r$_{xy}$) over a wide range of sheet current density (up to $\sim$1~A/m) and magnetic field (up to 9~T) provide a valuable insight into quantum Hall breakdown and a number of other non-linear phenomena (including electric instability, phase inversion of SdH oscillations, and a zero current anomaly). We have shown that these maps (phase diagrams) give detailed information about the conditions for, and the energetics of quantum Hall breakdown. Compared to earlier work in Ref.~\onlinecite{STUD2012}, we extended our study to the strong quantum Hall regime (reaching filling factor $\nu$=1). Also, an additional perspective was gained by incrementing the electron density in small steps by careful illumination and tracking the systematic evolution of the transport characteristics with n. We presented a simple picture for the principal features in the maps, namely the distinctive transport diamonds and high-current protrusions emerging from near the high current tips of the $\nu$=1 transport diamond, and introduced a simple tunneling model that qualitatively reproduces these features. Primary parameters such as critical current densities and critical Hall electric fields for $\nu$=1 and 2, and an exchange-enhanced g-factor for $\nu$=1 were extracted. The critical Hall electric fields for the InGaAs/InP quantum well were found to be comparable to those reported for widely studied GaAs/AlGaAs 2DEG hetero-structures. 

A detailed examination was made of the B-field dependence of the critical current I$_c$ as determined by two different methods and compiled for different values of n. From the first method (method A by which I$_c$ is defined to be the current when r$_{xx}$ exceeds a small threshold), we found I$_c$ for both even-$\nu$ and odd-$\nu$ has an approximate linear dependence with B-field. This finding is similar to that for an investigation in another In-based quantum well system \cite{ALEXANDAR2012}, but does not fit with the often cited B$^{3/2}$ dependence for the critical Hall field E$_c$. From the second method (method B by which I$_c$ is defined from the current position of the r$_{xx}$ maxima at the tips of the transport diamonds), we found I$_c$ for both even-$\nu$ and odd-$\nu$ also has an approximate linear dependence with B-field. While method B was easier to apply there was (on initial inspection) seemingly more scatter in data points. However, by simple rescaling, I$_c$/n data points for even-$\nu$ were found to collapse on to a single curve. This universal behavior uncovered on rescaling even-$\nu$ data points also implies that E$_c$ has a B$^2$ dependence when method B is applied. I$_c$/n data points for odd-$\nu$ do not all collapse onto a single curve and we attributed this to signs that the exchange-enhanced g-factor g$_o^*$ (a quantity that can be estimated from the slopes of the I$_c$/n traces for even- and odd-$\nu$) increases with decreasing n. This trend was corroborated by examination of the effective energy gap for odd-$\nu$ determined from conventional thermal activation measurements. By method B we estimated $\mid$g$_o^*{\mid}$ to be in the range $\sim$15-20 at base temperature. 

We have demonstrated that building up a map of quantum Hall breakdown provides access to numerous non-linear phenomena for carriers confined in two-dimensions that are driven out of equilibrium. The measurement and analytical techniques we have described here can in principle be applied to any material hosting a two-dimensional electron or hole system. Furthermore, method B can complement existing techniques such as the coincidence technique~\cite{NICHOLAS1988} and the thermal activation technique for estimation of exchange enhancement of spin splitting.

\section*{ACKNOWLEDGMENTS}

Part of this work is supported by NSERC (Discovery grant 208201) and support from RQMP and INTRIQ is acknowledged.

\appendix*
\section*{APPENDIX: TUNNELING DENSITY-OF-STATES MODEL} 

\begin{figure}[h!]
	\centering
	\hspace*{0cm}\includegraphics[width=0.4\textwidth]{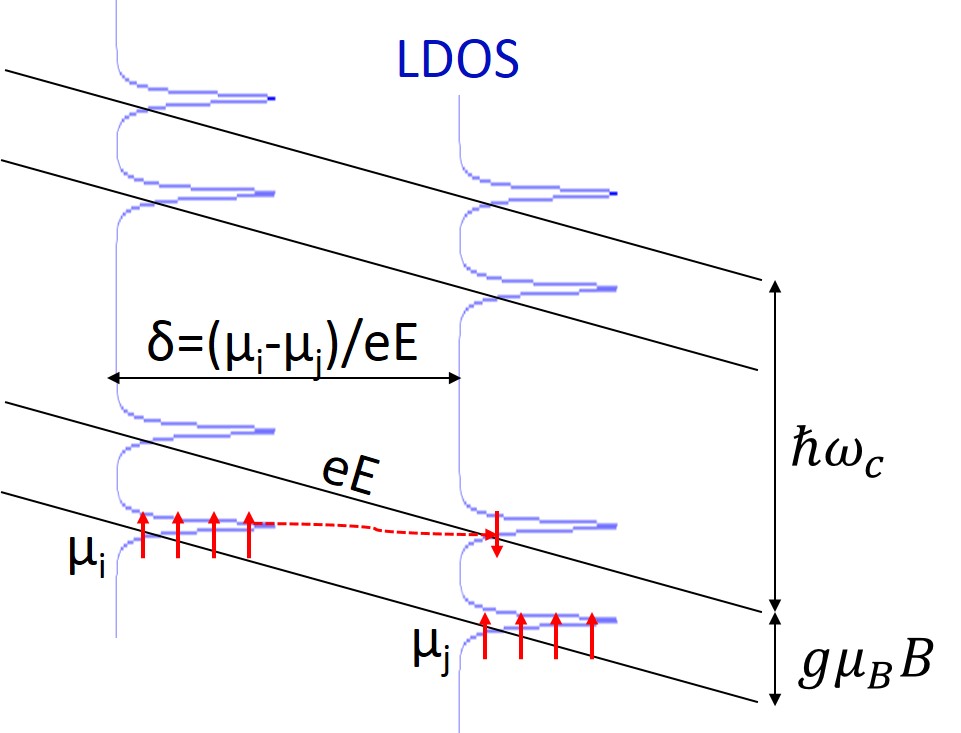}
	\caption{The local density-of-states (LDOS) including disorder broadening in a quantizing magnetic field is shown in blue. A sufficently large electric field $E$ (typically the transverse Hall electric field across the Hall bar) will lead to a Zener tunneling transition between regions $i$ and $j$ separated by distance $\delta$. Relevant for breakdown at $\nu$=1, the dashed red arrow depicts the transition (0,$\uparrow$)$\rightarrow$(0,$\downarrow$). The corresponding Landau levels are shifted in energy by $\mu_i-\mu_j$.}
	\label{fig:DOS1}
\end{figure}

\label{AppendixB} 

\newcommand\bra[1]{\left\langle#1\right|}
\newcommand\ket[1]{\left|#1\right\rangle}


Here we formulate a single-particle model describing the inter-Landau level tunneling processes exemplified in Fig.~\ref{fig:DOS1}.

\begin{figure}[h!]
	\centering
	\hspace*{0cm}\includegraphics[width=0.5\textwidth]{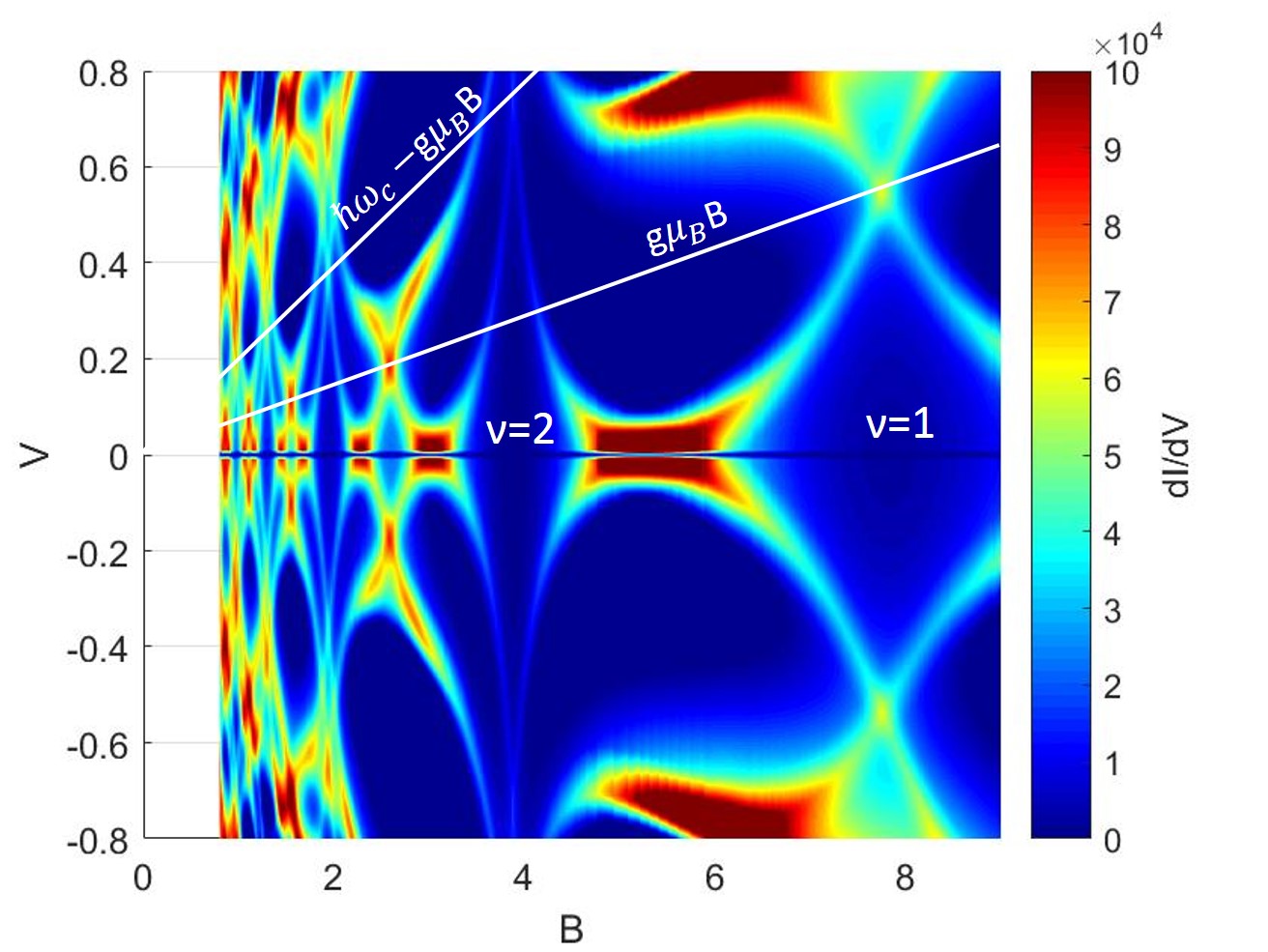}
	\caption{Example of a colormap of $\frac{dI}{dV}$ generated using expression (\ref{eq:B7}) with $B$ in units of $4\nu^{-1}n$, taking $n=1.9$ in units of $10^{11}$ cm$^{-2}$ (compare the colormap with the experimental colormap in Fig. 5). We have set $\hbar\omega_c/\textrm{g}\mu_B \textrm{B}=3.75$. This corresponds to g$\approx 11.5$: a value between g$_e^*$ and g$_o^*$  mentioned in section \ref{diamonds}. Since we have assumed a single value for the g-factor, we have neglected many-body interaction effects. We also used $\Gamma_{DOS}= \hbar\omega_c$ at 0.2~T, which is equivalent to $\sim$0.5~meV. $V$ and $dI/dV$ are in arbitrary units.}
	\label{fig:DOS2}
\end{figure}

We will assume that the system can be described by a simple tunneling Hamiltonian, which is given by the standard expression:
\begin{equation}
I=\frac{e}{h}\sum_{ij}T_{ij}n_i(1-n_j)\delta(\epsilon_j-\epsilon_i),
\end{equation}
where $T_{ij}$ are the the tunneling matrix elements. The local chemical potentials across the tunneling region are determined by the electric field $E$ and the distance $\delta$ between $i$ and $j$: $\mu_i=\mu+eE\delta/2$ and $\mu_j=\mu-eE\delta/2$. The uniform electric field in the model is governed by the DC bias, $V$, applied to the Hall bar (equivalent in the experiment to the DC current driven through the Hall bar which causes the Landau levels to tilt across the Hall bar). For simplicity we will assume $V=eE\delta$ and that the matrix elements $T_{ij}$ do not depend on energy, which leads at zero temperature to 

\begin{equation}
I(V)\sim\int_{\mu-eV/2}^{\mu+eV/2}d\epsilon D(\epsilon-eV/2)D(\epsilon+
eV/2),
\label{eq:B7}
\end{equation}
where $D(\epsilon)$ is the LDOS of the broadened Landau levels. The broadened Landau levels are typically rounded close to the Landau level energies (elliptic in the self-consistent Born approximation) with exponential tails between Landau levels. To mimic this behavior and in order to simulate the experimental findings, we assume the form of the Landau levels to be $e^{-|\epsilon/\Gamma_{DOS}|^{2/3}}$ in the tails and $e^{-|\epsilon/\Gamma_{DOS}|^{4/3}}$ close to the Landau levels ($\epsilon\simeq 0$), where $\epsilon$ is the energy relative to the center of a Landau level and $\Gamma_{DOS}$ is the broadening. The qualitative features will depend on the particular choice of Landau level broadening. We tried different forms but chose one that reproduces well the main features observed.

The differential conductance is given by g$\sim dI/dV$. To apply this model to the four-terminal experimental configuration, we note that for large magnetic fields where $\sigma_{xx}<\sigma_{xy}$, $r_{xx}\simeq\sigma_{xx}/\sigma_{xy}^2$: here $\sigma_{ij}$ are the magneto-conductivity tensor elements. Hence we assume $r_{xx}\sim dI/dV$, using expression \eqref{eq:B7}, which we evaluate as a function of $V$ and magnetic field. An example of a calculated r$_{xx}$ colormap is shown in Fig. \ref{fig:DOS2}. Although we do not expect our toy model to explain all aspects of the experimental r$_{xx}$ colormaps, it does nonethless reproduce a number of key features: (i) transport diamonds that grow in size with B-field and the alternating larger and smaller transport diamonds for even- and odd-filling factors; (ii) transport diamonds with clearly curved edges; and (iii) resonant features beyond the transport diamonds.

\bibliographystyle{apsrev4-1}
\bibliography{ref7}

\begin{thebibliography}{77}
\bibitem [{t()}]{t}%
  \BibitemOpen
  \href@noop {} {}\bibinfo {note} {For the fit calculation the Bohr radius,
  effective mass, and static dielectric constant are respectively set to 17 nm,
  0.047$m_0$ ($m_0$ is the free electron mass) and 15. The Si-doped InP layer
  above the QW is 5 nm wide and nominally doped to 6.7$\times 10^{17} cm^{-3}$
  so we set n$_i$=3.35$\times$10$^{11}$~cm$^{-2}$ where n$_i$ is the effective
  2D density of the remote ionized impurities (we assume the donors are fully
  ionized). The setback distance is set to 42.5 nm (this is the distance from
  the middle of the 10~nm wide In$_{0.76}$Ga$_{0.24}$As channel to the middle
  of the Si-doped InP layer). Fit parameters are
  $\mu_A$=349000$\pm$7000~cm$^2$/Vs, and
  n$_B$=(1.8$\pm$0.2)$\times$10$^{15}$~cm$^{-3}$ where n$_B$ is the effective
  density of the background ionized impurities. The value for n$_B$ is in line
  with expectations for the chemical beam epitaxy system used to grow the
  hetero-structure.}\BibitemShut {Stop}%
\bibitem [{Q()}]{Q}%
  \BibitemOpen
  \href@noop {} {}\bibinfo {note} {For method A, generally data points for
  even-$\nu$ could be extracted above $\sim$1.5~T, however, above 6~T
  determination for $\nu$=2 was not straightforward because of fluctuations or
  multiple-step transitions prior to abrupt breakdown when
  n$>$2.7$\times$10$^{11}$~cm$^{-2}$. For odd-$\nu$, other than for $\nu$=1
  which moved out of range when n$>$1.9$\times$10$^{11}$~cm$^{-2}$, only data
  points for $\nu$=3 could be extracted over a wide range of n. For method B,
  more data points could be extracted and this was generally straightforward
  above $\sim$1~T. However, for even-$\nu$ determination was not possible or
  straightforward if the transport diamond width was close to or exceeded the
  current sweep range (for example, $\nu$=2 at
  n=1.9$\times$10$^{11}$~cm$^{-2}$), or multiple-peak features developed at
  breakdown (for example, $\nu$=2 at n=2.3$\times$10$^{11}$~cm$^{-2}$: note the
  data points for this density fall short of the common lines in the lower two
  panels of Fig.~6 for even-filling factor points after rescaling), or
  widespread instability occurred beyond breakdown ($\nu$=2 beyond $\sim$5~T
  for n$>$2.3$\times$10$^{11}$~cm$^{-2}$).}\BibitemShut {Stop}%
\bibitem [{SCA()}]{SCATTER}%
  \BibitemOpen
  \href@noop {} {}\bibinfo {note} {In Figs.~6(c) and 6(d), even after
  rescaling, there is more scatter in odd-integer filling data points than for
  even-integer filling factor points. As discussed later in Sec. VII, not only
  can this be a sign of an exchange-enhanced g-factor that varies with n, but
  also of an exchange-enhanced g-factor that varies with B-field at constant
  n.}\BibitemShut {Stop}%
\bibitem [{AAA()}]{AAA}%
  \BibitemOpen
  \href@noop {} {}\bibinfo {note} {For n=1.9$\times$10$^{11}$~cm$^{-2}$ the
  estimate of $\mid$g$_o^*{\mid}$ for $\nu$=1 from the slopes of the lines in
  Fig.~6(c) is higher than the crude estimate of $\mid$g$_o^*{\mid}$ for
  $\nu$=1 given in Sec. VI. For that estimate, however, Landau level broadening
  was neglected and so the $\mid$g$_o^*{\mid}\mu_B$B line bounding odd-$\nu$
  diamonds was assumed to extrapolate to the origin (see Fig.~5). However,
  Fig.~6(c) reveals odd-$\nu$ data points generally lie on lines that do not
  extrapolate to the origin but instead intercept zero current near 1~T. As a
  consequence, the slope and hence the extracted g-factor will be
  larger.}\BibitemShut {Stop}%
\bibitem [{y()}]{y}%
  \BibitemOpen
  \href@noop {} {}\bibinfo {note} {The FWHM value of $\sim$0.4~$\mu$A is a
  typical value for the selected sections shown in Fig.~8. Examination of other
  sections near where neighboring transport diamonds touch give values for the
  FWHM as low as 0.2~$\mu$A.}\BibitemShut {Stop}%
\end{thebibliography}

\end{document}